# Risk-based framework to determine climate-informed design storms for road drainage infrastructure

Mohammad Fereshtehpour[1*] , Rashid Bashir[1*], Neil F. Tandon[2, 3]

[1] Department of Civil Engineering, Lassonde School of Engineering, York University, Toronto, ON, Canada
[2] Department of Earth and Space Science and Engineering, York University, Toronto, ON, Canada
[3] Centre for Research in Earth and Space Science, York University, Toronto, ON, Canada

* Corresponding author: mferesht@uwo.ca

**Abstract**

Climate change is amplifying extreme precipitation events in many regions and imposes substantial challenges for the resilience of road drainage infrastructure. Conventional design storm methodologies, which rely on historical trends of rainfall data under a stationarity assumption, may not adequately account for future climate variability. This study introduces a risk-based framework for determining climate-informed design storms tailored to road drainage systems. The proposed framework integrates climate model projections with risk assessment to quantify the potential impacts of future extreme rainfall on drainage performance and adjust the future design storm, with a focus on the province of Ontario, Canada. Projected precipitation changes for mid- and late-century time horizons are quantified using statistically downscaled CMIP6 General Circulation Models (GCMs). The risk level is defined as a function of hazard and vulnerability, where hazard combines both physiographic and meteorological factors. Vulnerability is comprised of socioeconomic, transportation, and environmental considerations. To systematically integrate these components, a weighting scheme is developed based on a sensitivity analysis of the criteria, which provides flexibility in assigning relative importance to each factor. The estimated risk level is then applied to adjust projected design storm accordingly. The proposed workflow is demonstrated through both province-wide and site-specific applications across Ontario's road network to better highlight its scalability and adaptability. The findings signify the necessity of shifting from static, stationarity-based design methodologies to dynamic, risk-informed approaches that enhance the long-term resilience of transportation networks.

**1. Introduction**

Well-planned and resilient infrastructure serves as the backbone of societal stability and productivity. Due to the exposure of critical infrastructure to natural hazards and their long service life and substantial capital investment, infrastructure systems are particularly vulnerable to failure, which can result in far-reaching economic and social disruptions (Argyroudis et al., 2020). Climate change is expected to intensify these risks—under a high-emissions scenario, nearly 44% of global transport assets could face a 25% reduction in extreme rainfall design return period by mid-century, rising to ~70% by late-century as warming approaches 4°C (Liu et al., 2023). This underscores the importance of enhancing their long-term resilience (Islam et al., 2024). Infrastructure resilience refers to the capacity of these systems to endure, adapt to, and recover from a wide range of disturbances while maintaining or improving their functionality (Tabasi et al., 2025). Designing infrastructure with resilience in mind from the outset is essential to reducing future environmental and financial risks (Shadabfar et al., 2022). Evidence suggests that investing just 2% of road value in improved drainage and flood protection can generate positive returns for nearly 60% of roads that face at least one flooding event (Koks et al., 2019).

Highways are critical transportation infrastructure facilitating the movement of goods and people across regions. Flooding can cause roads to become blocked or partially underwater, which creates major obstacles that prevent people from evacuating safely (Seng et al., 2024). The appropriate design of highway drainage is critical to not only preventing flooding during normal and extreme conditions but is integral to proper road and highway maintenance. Their operational importance necessitates robust design and construction practices to mitigate the impact of diverse meteorological and climatic stressors. Flooding poses a major hazard to highways, disrupting traffic, damaging infrastructure, and hindering regional economic activity (Pedrozo-Acuña et al., 2017). High-water levels degrade road infrastructure performance and longevity by increasing incidents like landslides, road washouts, bridge support submersion, and road closures (Ou-Yang et al., 2015). Beyond safety concerns during flood events, maintaining the functionality of critical roadways, particularly those needed for emergency services, is essential. These roads form the backbone of urban infrastructure, but they can be highly vulnerable to flooding during extreme storms (Pregnolato et al., 2017).

Conventional practice designs drainage systems by analyzing long-term rainfall records to derive representative design storm events. However, many drainage structures, including culverts, trenches, and bridges, are not adequately equipped to handle the current frequency of extreme flow events. For instance, in 2015, South Carolina experienced a historic 1000-year rainfall event (NOAA, 2016), leading to catastrophic flooding and widespread road infrastructure failure. Nearly 400 roads were initially closed, largely due to washouts caused by the collapse of pipe culvert soil systems (Gassman et al., 2017). The failure mechanism, while uncertain, likely originated from either structural overload of the culvert due to hydraulic capacity exceedance or erosion-induced loss of soil support at the inlet/outlet, ultimately compromising structural integrity (French and Jones, 2015; Gassman et al., 2017). Accelerating climate change and urbanization are intensifying the frequency and severity of extreme rainfall events, making traditional methodologies inadequate (Shao et al., 2024). Projected increases in extreme hydro-meteorological event frequency and intensity are expected to impact road infrastructure design parameters and lifecycle costs (Pedrozo-Acuña et al., 2017).

Drainage infrastructure is typically designed using Intensity-Duration-Frequency (IDF) curves, which characterize the relationship between rainfall intensity, storm duration, and the likelihood of occurrence for a specified return period (Schardong et al., 2020). The design storm is an important output from the IDF curves which refers to a hypothetical storm used in hydrological modeling to predict rainfall and estimate potential flooding or drainage impacts. Design standards for drainage infrastructure worldwide largely rely on the assumption of stationarity, using historical rainfall data to develop IDF curves (Underwood et al., 2020). In the United States, for example, stationary IDF curves from NOAA Atlas 14 (NA14) are widely used as the benchmark for stormwater infrastructure design (Ghasemi Tousi et al., 2021). Drainage infrastructure designed based on stationary IDF curves may be inadequately engineered and increasingly vulnerable to the impacts of climate change (Lopez-Cantu and Samaras, 2018).

A growing body of evidence shows that climate change is inherently non-stationary, reshaping rainfall regimes and intensifying extreme events, thereby compelling engineers to update and adapt conventional design methodologies (Mallakpour and Villarini, 2015; IPCC, 2023). The 2022 IPCC reports attribute the increasing frequency of extreme storms to global warming, projecting a 7% rise in rainfall intensity for each degree of temperature increase (IPCC, 2023). But there are strong regional variations in these projections due to localized changes in atmospheric motions (Tandon et al., 2018), necessitating a more regionally aware approach to risk assessment. Given the growing impacts of climate change on stormwater drainage and flooding, a thorough re-evaluation of existing design practices is essential to identify and implement adaptive solutions (Haghighatafshar et al., 2020). Design standards are increasingly adapting to climate change by updating IDF curves with projections from General Circulation Models (GCMs) (Kourtis and Tsihrintzis, 2022). Despite these efforts, incorporating GCM output into IDF curves introduces significant uncertainty (Fadhel et al., 2021). These uncertainties largely stem from the use of different GCMs, structural differences among models, the choice of greenhouse gas emission scenarios, differences in projected time horizons, and the application of various downscaling and bias correction techniques (Underwood et al., 2020; Fereshtehpour and Najafi, 2025). The choice of climate model and future scenario significantly impacts estimates of extreme events. For example, in a stormwater culvert design, different model outputs could lead to a doubling of the required size, substantially increasing costs (Ghasemi Tousi et al., 2021). This highlights the need for a framework that effectively integrates climate projections while addressing associated uncertainties to improve decision-making processes.

Risk-based design is indispensable at the intersection of climate-change adaptation, urban planning, infrastructure systems, and social dynamics, where these factors collectively determine a community's flood resilience (Haghighatafshar et al., 2020). Flood risk is typically determined by both flood probability and its consequences and a risk-based paradigm expands cost-effective solutions compared to traditional approaches, which offers policymakers greater flexibility to maintain or improve service levels at the same or lower cost (Roth, 2014). Transitioning from a probability-based system centered on recurrence intervals to a risk-based approach incorporating probabilistic assessments of flood likelihood and consequences is crucial for managing climate change uncertainties, urbanization, and population growth (Haghighatafshar et al., 2020; Kharoubi et al., 2023). For instance, unlike traditional approaches that typically design for a 100-year flood event, a risk-based approach allows for a more flexible protection target, which may be either lower or higher than the 100-year flood, depending on the flood probability, potential consequences, and associated infrastructure costs (Rosner et al., 2014).

Current approaches to the design and planning of transportation hydraulic infrastructure often rely on uniform standards that do not fully account for the spatial variability of hazard exposure or the potential consequences of failure. These methods typically emphasize infrastructure type when specifying design parameters, with limited consideration of regional context, system vulnerability, or climate change uncertainty (MTO, 2023a). Road networks span large, diverse geographies and are governed by different agencies that require approaches that are standardized but also flexible in accommodating local perspectives. In regions with low accessibility and high environmental or socio-economic stakes, the consequences of infrastructure failure can be particularly severe. These realities underscore the need for a comprehensive and spatially scalable risk assessment framework that incorporates transportation system vulnerability and supports proactive adaptation across entire networks. While existing frameworks (Wall and Meyer, 2013; Toplis, 2015) excel at defining high-level governance structures, qualitative risk registers, and policy-driven scenario planning, they rely on historical design-storm approaches and narrative scenarios rather than prescribing specific methods to translate climate-model projections into updated design storms for drainage assets.

The primary obstacle for the field lies in the absence of a standardized methodology for integrating climate change considerations into the design of highway infrastructure. This paper addresses this gap by presenting a novel risk-based rainfall adjustment framework. This research marks a significant advancement as it is the first attempt of this type to develop such a framework specifically for highway drainage infrastructure. The proposed framework integrates multiple hazard and vulnerability components to inform climate-resilient infrastructure planning. Its key contributions include (a) enabling engineers and policymakers to develop more risk-informed highway drainage systems, (b) preventing over- or under-design by considering specific hazard exposures and intrinsic vulnerabilities, (c) supporting a shift from traditional stationary design methods to dynamic approaches that account for future climate uncertainties, and (d) providing a comprehensive and spatially scalable risk assessment framework that incorporates transportation system vulnerability and supports proactive adaptation across entire networks. The proposed framework aims to quantify anticipated changes in future rainfall and integrate these into design practices. Shifting from a static, trend-based approach to a dynamic adjustment offers a more cost-effective and adaptable way to manage climate uncertainties and enhance highway infrastructure resilience. While frameworks for climate adaptation in road infrastructure are available, they do not offer this specific type of detailed, rainfall-adjustment focused methodology for highway drainage.

This paper is structured to first outline the materials and methods, detailing a five-step process for implementing the risk-based adjustment framework. This includes defining risk components, integrating climate projections, structuring a hierarchical assessment, applying a weighting scheme based on a sensitivity analysis of the criteria, and adjusting future rainfall projections based on risk levels. Subsequently, we present results and discussion regarding applications of the framework to three potential locations in Ontario, analyzing risk assessment outcomes, spatial distribution of risk levels, and the implications of these risk-based rainfall adjustments. Finally, the paper concludes by emphasizing the necessity of a non-stationary, risk-informed approach for resilient infrastructure planning.

## 2. Study Area

Our study area is Ontario, Canada's most populous province, which features a diverse landscape with extensive transportation networks that are critical for economic activity and public mobility (Kennedy et al., 2009). The study area includes major urban centers such as Toronto and Ottawa, and key transportation corridors like Highway 401, one of North America's busiest highways. The study area also includes remote communities in Northern Ontario, many of which rely on a single road that is increasingly threatened by climate change, affecting access to essential goods and services (Douglas and Pearson, 2022; Hori, 2016). This study focuses on assessing risk for highway drainage infrastructure within the Province of Ontario, covering all of the five administrative regions defined by the Ministry of Transportation of Ontario (MTO) (Figure 1). The foundational principles of the proposed methodology are designed for broad, cross-contextual relevance. Although demonstrated within a regional scope, its inherent transferability and scalability facilitate its effective application in diverse geographic settings, provided that analogous data are available.

The proposed workflow is implemented at two spatial scales: a province-wide analysis and site-specific evaluations. At the provincial scale, the framework leverages the Ontario Road Network (ORN)—the authoritative geospatial dataset maintained by the Ontario Ministry of Natural Resources (OMNR, 2025), which encompasses municipal roads, provincial highways, and resource and recreational routes. This broad-scale application highlights the scalability of the approach and its utility in supporting regional infrastructure planning. At the local scale, the framework is applied to culvert-level assessments at three representative road segments, each located in a different MTO region, demonstrating its adaptability for detailed, site-specific decision-making. These include South Malden Road, a local municipal road in the Western region crossing the Conard River, Finch Avenue, an arterial road in the Central region crossing Black Creek, and Highway 17, an expressway/highway in the Northeastern region crossing Pickerel Creek. The selected locations, shown in Table 1 and Figure S1-S3, are chosen to capture a range of road classifications and regional hydrologic contexts, further illustrating the adaptability of our methodology to varying infrastructure and environmental settings.

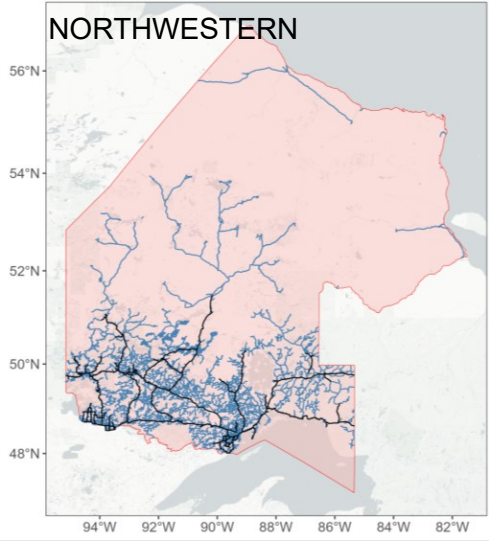
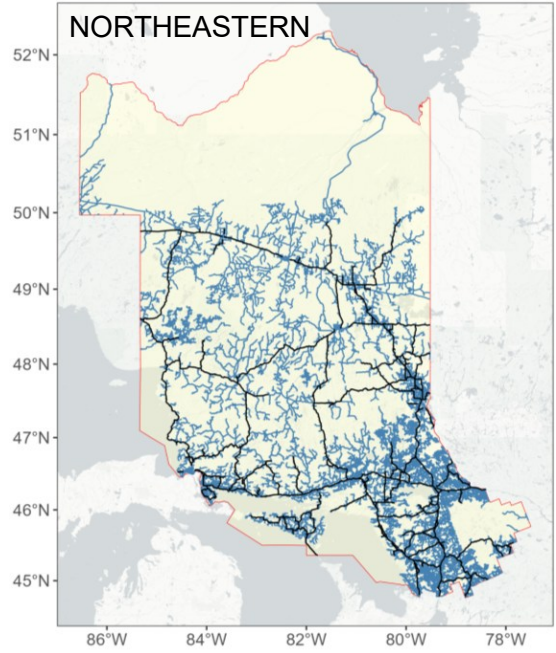
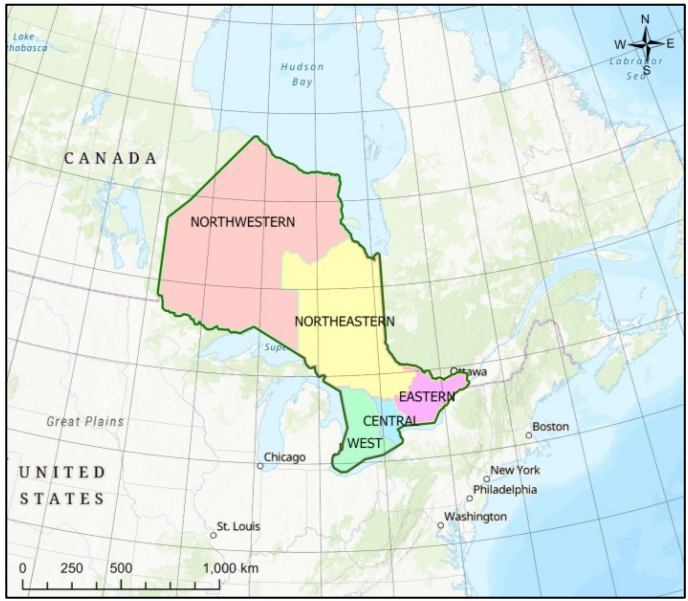
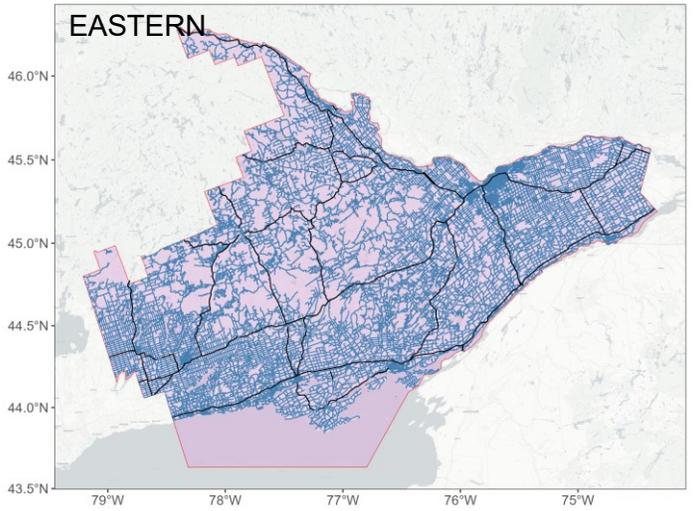
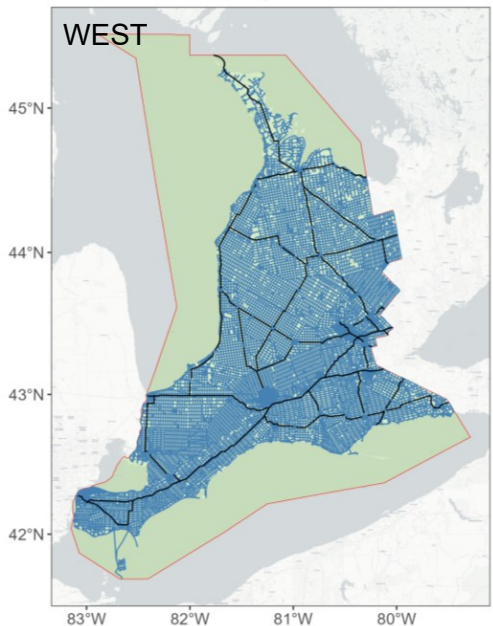
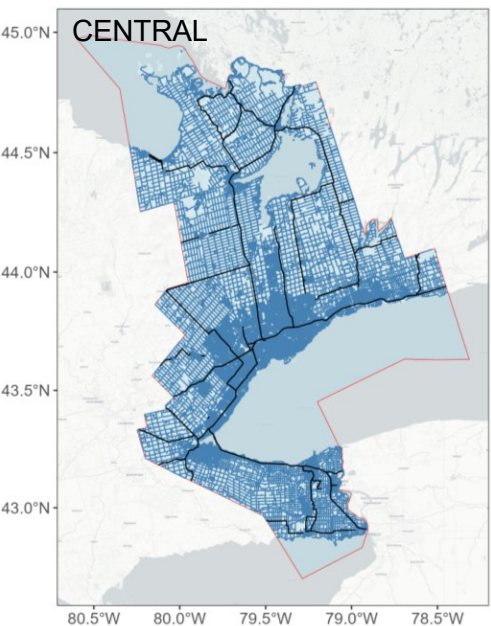

**Figure 1.** Boundaries of the five MTO administrative regions and the road networks in each region across Ontario, Canada. Black line represents the highway networks.

**Table 1.** Summary of selected road segments in Ontario used for site-specific assessments of drainage infrastructure (culverts).

| Road name | Road type* | Region** | River traversed | Latitude, longitude |
|---|---|---|---|---|
| South Malden Rd. | Local Municipal Road | West | Conard River | 42.1233°N, 82.8482°W |
| Finch Ave. | Arterial | Central | Black Creek | 43.7600°N, 79.5048°W |
| Highway 17 | Local Arterial (Expressway/Highway) | Northeastern | Pickerel Creek | 46.2878°N, 83.4035°W |

** Road types are based on the Geometric Design Guide (GDG) for Canadian Roads (TAC, 2017)
* According to the boundaries of the MTO administrative regions.

## 3. Risk-based methodology

The proposed workflow, depicted in Figure 2, employs a systematic multi-criteria decision-making (MCDM) framework, currently utilizing a sensitivity-based weighting approach to quantify the relative significance of various risk factors. Hazard levels are derived through the evaluation of key physiographic and meteorological variables, while vulnerability is assessed by synthesizing socio-economic, transportation, and environmental indicators. To ensure methodological consistency and robustness in risk assessment, the approach incorporates both quantitative geospatial data (e.g., hazard maps) and qualitative condition-level classifications. The resultant composite risk index, evaluated at specific infrastructure locations, informs the dynamic adjustment of design storm intensities under future climate change scenarios—specifically, the SSP2-4.5 (intermediate) and SSP5-8.5 (high) greenhouse gas concentration pathways. The maximum projected percentage increase in precipitation intensity across scenarios and time horizons is utilized to linearly scale the baseline design storm associated with a given return period.

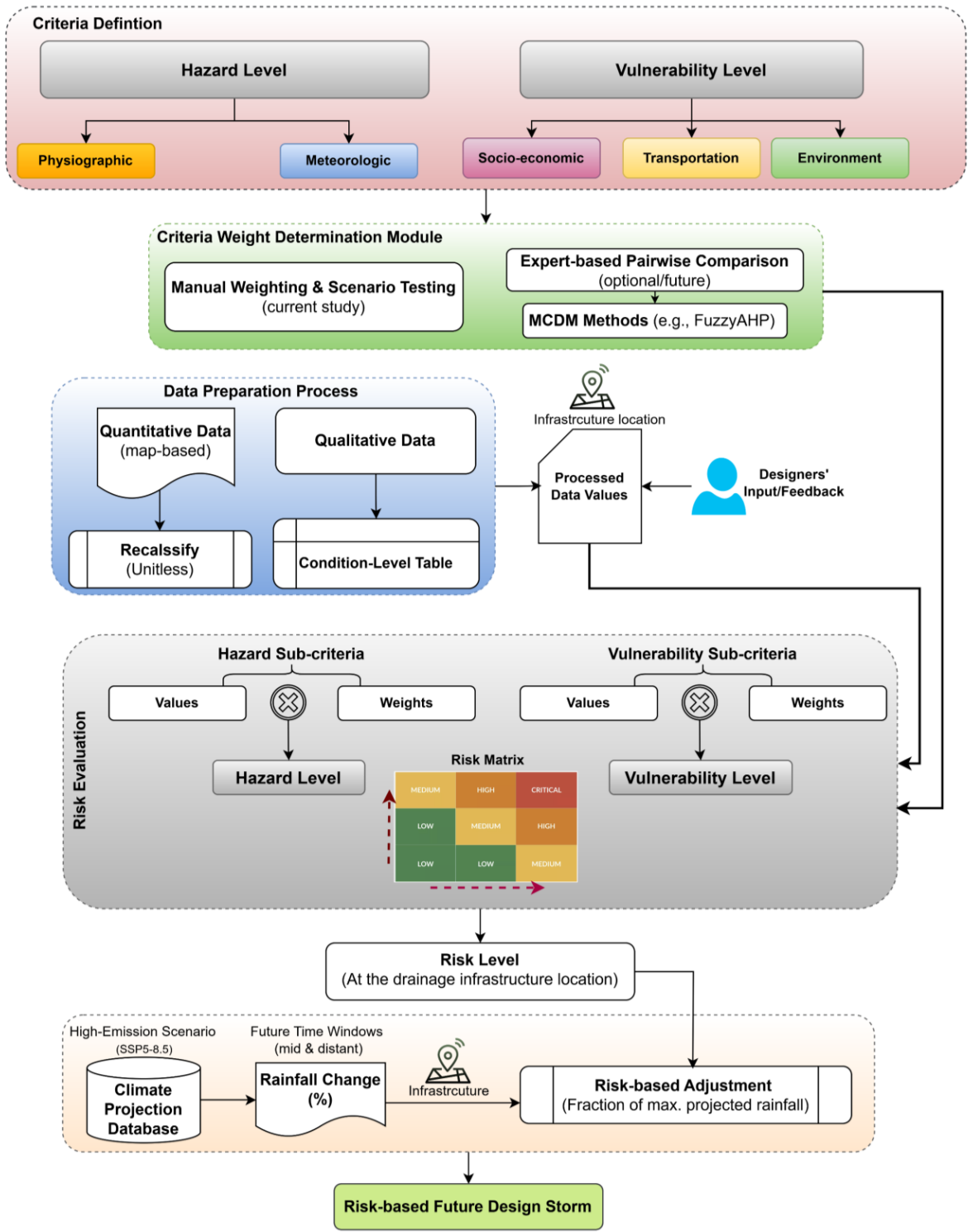

**Figure 2.** Proposed workflow for the risk-based future design storm.

## 3.1. Risk assessment framework

Risk is fundamentally a combination of hazard and vulnerability, where hazard represents the likelihood and magnitude of a natural event, and vulnerability reflects the susceptibility of exposed systems to the natural event's impacts (Tabasi et al., 2025). Areas with high flood hazards may not experience severe consequences if vulnerability is low. Conversely, highly vulnerable regions can suffer significant impacts even from moderate flooding events. Risk is also affected by the ability of human and environmental systems to withstand, respond to, and recover from these events (Shao et al., 2024). Therefore, the relationship between hazard and vulnerability is dynamic, meaning that risk varies across spatial and temporal scales, influenced by both natural and human-driven processes. To this end, the first step of determining risk in this context is defining the key criteria to capture both the likelihood of flooding events and the potential consequences on transportation systems and surrounding communities. Figure 3 illustrates the proposed hierarchical structure used to define risk levels, incorporating all criteria and sub-criteria across multiple levels of the hierarchy.

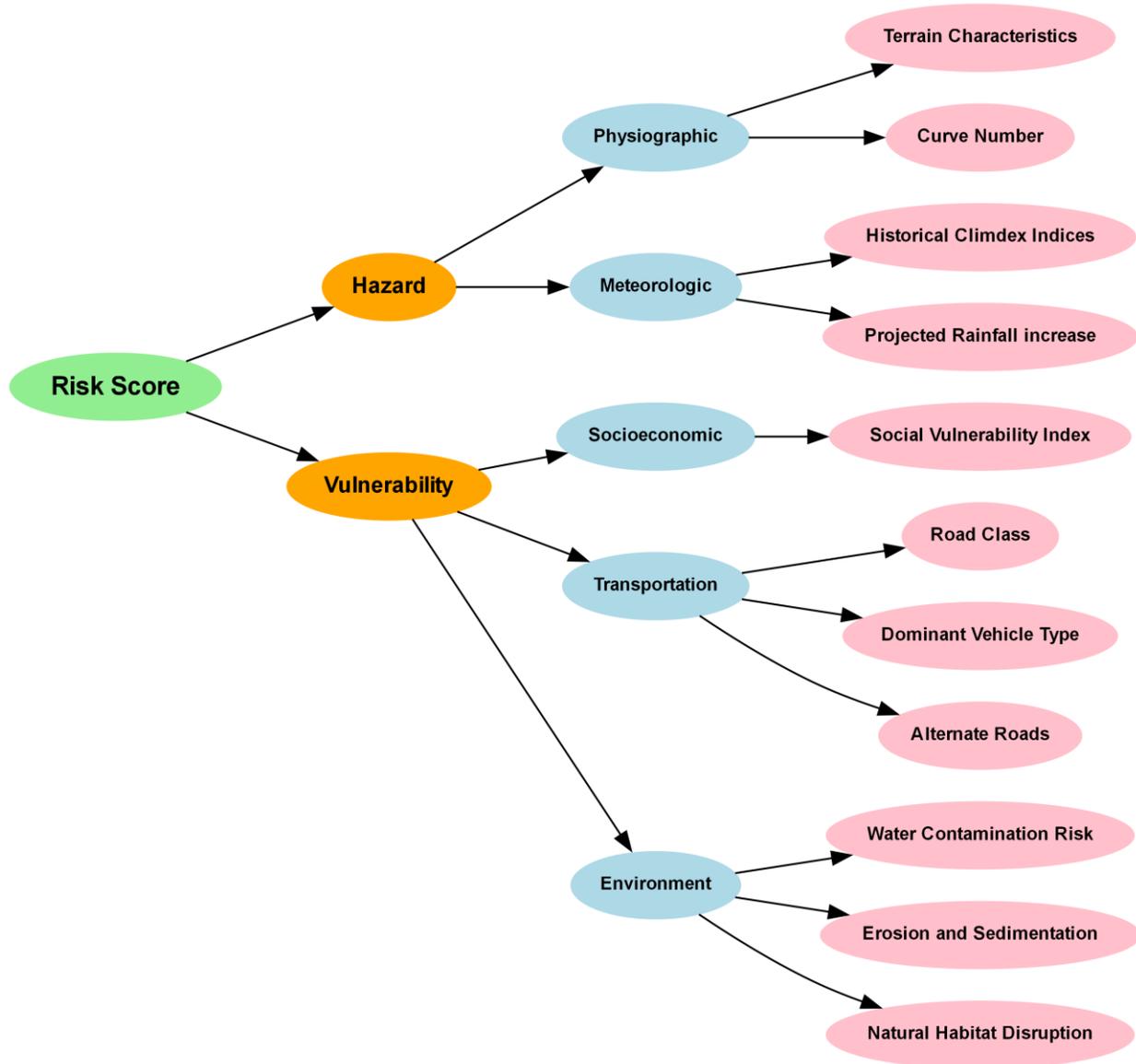

**Figure 3.** Hierarchical structure of the flood risk components.

### 3.1.1. Hazard Level

Hazard level maps for flood risk estimation are developed by integrating physiographic and meteorological data to quantify spatial flood hazard variations. The physiographic component includes terrain characteristics, which influence surface runoff dynamics, along with the Curve Number (CN) which represents infiltration and runoff potential. The meteorological component incorporates climate extreme indices, such as extreme precipitation percentiles, to assess the frequency and intensity of extreme rainfall events. Additionally, projected future rainfall increases from climate model output are incorporated to account for potential changes in flood hazard under different climate scenarios. The combined analysis produces hazard level map that delineate flood-prone areas.

As mentioned above, physiographic indicators include (1) terrain characteristics and (2) land cover-based runoff potential. The terrain features are derived from the Ontario Provincial DEM (PDEM) of 30-meter spatial resolution (OMNRF, 2019). Figure 4 provides a continuous representation of elevation across the study area. The color gradient indicates varying elevations, where lower elevations are primarily associated with lake basins, river valleys, and low-lying terrain, while higher elevations correspond to elevated landforms such as the Canadian Shield, highlands and escarpments. In this paper, several key features are extracted from the PDEM to assess hydrological characteristics that influence water retention, runoff patterns, and flood susceptibility. These features include slope, flow accumulation, and the topographic wetness index (TWI), each of which contributes significantly to identifying flood-prone areas and assessing drainage efficiency. Slope represents the rate of elevation change in a given area and is a key factor in determining runoff velocity and erosion potential (Dutta and Deka, 2024). Steeper slopes lead to faster runoff and reduced infiltration, while gentler slopes allow for greater water retention and potential ponding. Flow accumulation is determined by counting the total number of pixels that naturally contribute drainage to downstream outlets. It is derived from a flow direction model that routes water based on the terrain gradient. Higher flow accumulation values indicate areas where water is likely to converge and form channels or rivers (Howlader et al., 2024). The TWI is a measure of soil moisture potential based on terrain characteristics (Riihimäki et al., 2021). It quantifies how likely a location is to retain water based on its slope and upslope contributing area.

TWI is computed as

$$TWI = \ln\left(\frac{A}{\tan S}\right) \quad (1)$$

where $A$ is the specific catchment area (SCA, i.e., the flow accumulation per unit contour length) and $S$ is the slope in radians.

These three factors allow for a balanced assessment of flood hazards, particularly in relation to road drainage networks. In this context, other factors, such as distance from rivers, contribute little to the variability in flood potential, as most drainage systems are constructed in close proximity to river systems. A steep slope facilitates rapid runoff, reducing water retention and minimizing infiltration, whereas areas with high flow accumulation act as natural drainage pathways where water converges (Dutta and Deka, 2024). However, flow accumulation alone does not determine whether an area retains water or drains efficiently—this is where TWI plays a crucial role (Abdelkareem and Mansour, 2023). While high flow accumulation indicates major drainage paths, TWI distinguishes between well-drained areas and those prone to water retention based on slope and upstream contributing area. Steep slopes generally correspond to low TWI values, signifying drier conditions due to rapid runoff, whereas gentle slopes with high flow accumulation result in high TWI, indicating areas prone to waterlogging and sustained moisture retention. These relationships highlight the necessity of considering all three factors together for more realistic and infrastructure-oriented flood risk assessments.

In addition to terrain characteristics, soil properties and permeability conditions significantly influence runoff potential (Tabasi et al., 2025). In this study, we adopt curve number (CN), which is a widely used hydrological model parameter for estimating direct runoff, based on land use, soil type, and antecedent moisture conditions. The CN values directly influence runoff depth estimation, which in turn affects flood potential. Higher CN values (>90) indicate low infiltration and high runoff potential, typical of impervious or compacted surfaces, thus increasing flood risk. Lower CN values (<50) reflect higher infiltration and reduced runoff, typical of permeable, well-vegetated soils (USDA, 1986). The CN varies depending on the soil's initial moisture state, which can be classified into three categories: dry (CNI), average (CNII), and wet (CNIII) conditions (Soulis and Valiantzas, 2012). The dry condition represents a scenario where the soil has low antecedent moisture, resulting in higher infiltration capacity and reduced runoff generation. Conversely, the wet condition corresponds to saturated soil with minimal infiltration capacity, leading to higher runoff potential. Our analysis focuses on the average CN as the principal parameter for runoff estimation, while dry and wet CN values are considered for supplementary assessment to account for variations in antecedent soil moisture. A globally consistent, high-resolution (250 m) gridded dataset of CN values, developed by Jaafar et al. (2019), is employed to provide spatially detailed and standardized hydrological inputs across the study area (Figure 4 e-f).

It is important to note that, according to the MTO Hydrotechnical Design Chart 1.09, CN values for natural areas should be derived based on the Hydrologic Soil Group (HSG), which categorizes soils based on their infiltration characteristics (MTO, 2023b). In urban environments, impervious surfaces drastically influence runoff characteristics. The highest CN values are associated with fully paved areas such as roofs and parking lots, where infiltration is negligible, and nearly all precipitation is converted into direct runoff. These surfaces are assigned a CN value of 98, indicating near-total runoff production (Cronshey, 1986). This distinction between natural and urbanized landscapes is needed for flood modeling and assessment, as areas dominated by impervious surfaces contribute more significantly to peak discharge and flood hazard, particularly in high-intensity rainfall events.

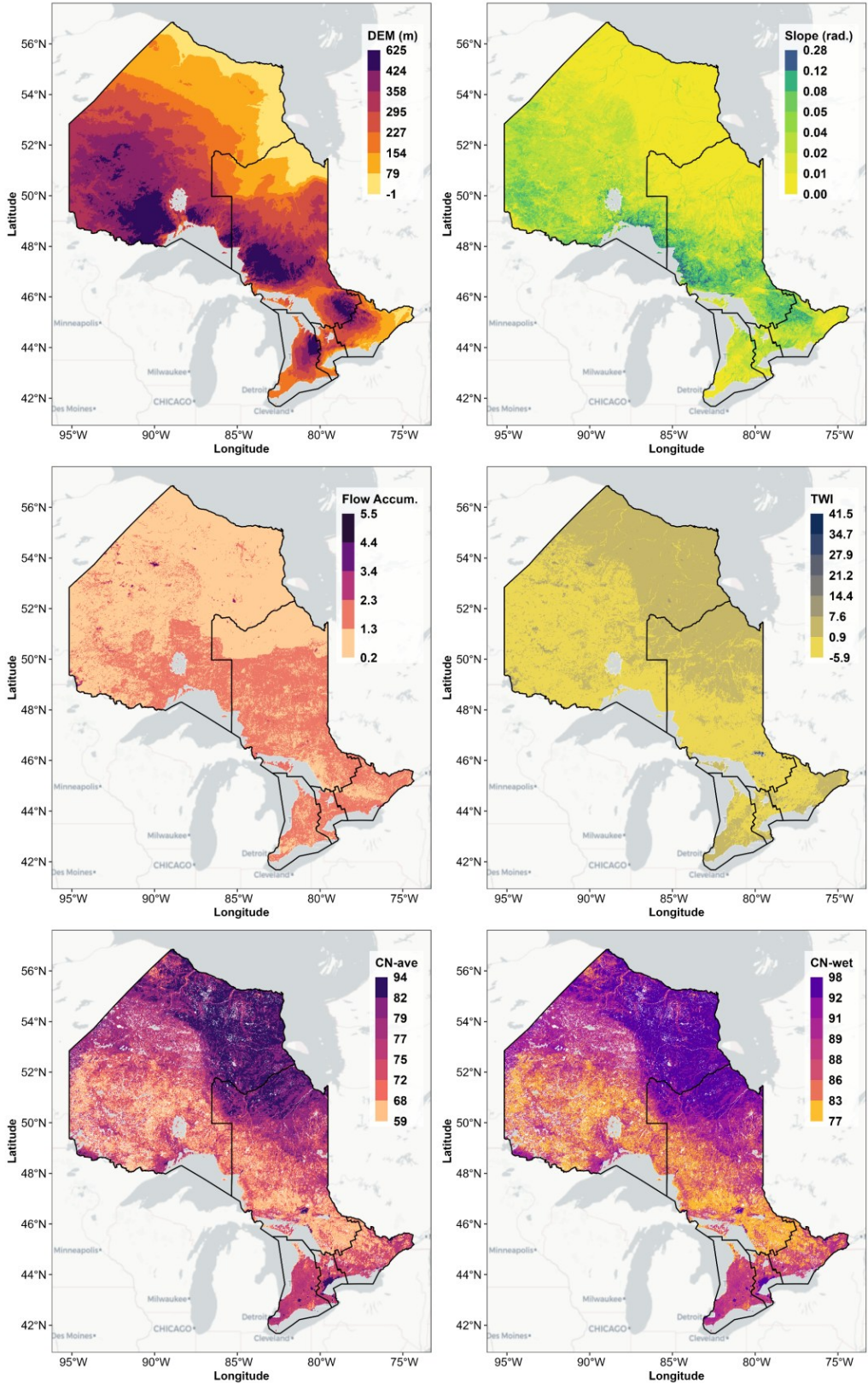

**Figure 4.** Digital Elevation Model (DEM) of Ontario with 30-meter resolution (OMNRF, 2019) and derived physiographic indicators: (b) Slope (radians), (c) Flow Accumulation, and (d) Topographic Wetness Index (TWI). The DEM is available from the Ontario GeoHub (https://geohub.lio.gov.on.ca/). Panels (e) and (f) illustrate the average and wet-condition Curve Number (CN) values, respectively (original resolution: 250 m, resampled to 1 km).

Meteorological factors were incorporated into the analysis through two key components: (1) a composite indicator based on the selected "Climdex indices" representing precipitation-related climate extremes, and (2) projected changes in future extreme precipitation derived from climate model output. The Climdex indices are internationally recognized metrics to quantify changes in the frequency, intensity, and duration of extreme climate events. Detailed descriptions of these indices are available at https://www.climdex.org, as recommended by the Expert Team on Climate Change Detection and Indices (ETCCDI). In this study, six precipitation-related indices— PRCPTOT, R99, R95, R20mm, SDII, and CWD—were initially considered. Due to the observed redundancy and strong inter-correlations among certain indices, three representative indicators (PRCPTOT, R99, and SDII) are ultimately selected (Figure 5). These indices provide complementary characteristics in representing both total and extreme precipitation behavior under current climate scenarios. The definitions and mathematical formulations of these indices as follows. The PRCPTOT index (annual total wet-day precipitation) quantifies the total accumulated precipitation from wet days, defined as days with daily precipitation exceeding 1 mm. It is mathematically defined as:

$$PRCPTOT = \sum_{i=1}^{n} P_i \quad P_i > 1 \text{ mm}, \quad (2)$$

where $n$ is the total number of wet days in a year. Although it is influenced by both the frequency and intensity of wet events, it does not differentiate between them, so changes in this index alone do not indicate which of these components is dominant.

The R99p index (extremely wet days) represents the annual total precipitation from days where rainfall exceeds the 99th percentile of the daily precipitation distribution during a reference period. It is given by:

$$R_{q_{th}} = \sum_{i=1}^{n} P_i \qquad P_i > P_{q_{th}}, \tag{3}$$

where $q_{th}$ refers to the 99th percentile value of precipitation over the wet days of the historical period and $n$ is the total number of wet days in a year given the specified condition.

Finally, the SDII (Simple Daily Intensity Index) captures the average intensity of precipitation on wet days. It is calculated as the ratio of total precipitation to the number of wet days:

$$SDII = \frac{\sum_{i=1}^{n} P_i}{n} \qquad P_i > 1mm \tag{4}$$

In terms of hydrological impacts, increases in PRCPTOT suggest a rise in overall precipitation volumes, which can elevate baseflow and cumulative runoff. In contrast, increases in R99p or SDII reflect more intense rainfall events, which are strongly correlated with increased surface runoff, higher peak flows, and an elevated risk of flash flooding, particularly in urban and poorly drained areas. In this paper, the historical indices are calculated based on CMORPH data (Joyce et al., 2004) over a 26-year period from 1998 to 2023.. CMORPH produces rainfall estimates globally at very high spatial (~8 km) and temporal (30 min) resolutions. Figure 5 shows the mean annual precipitation across Ontario using this CMORPH dataset.

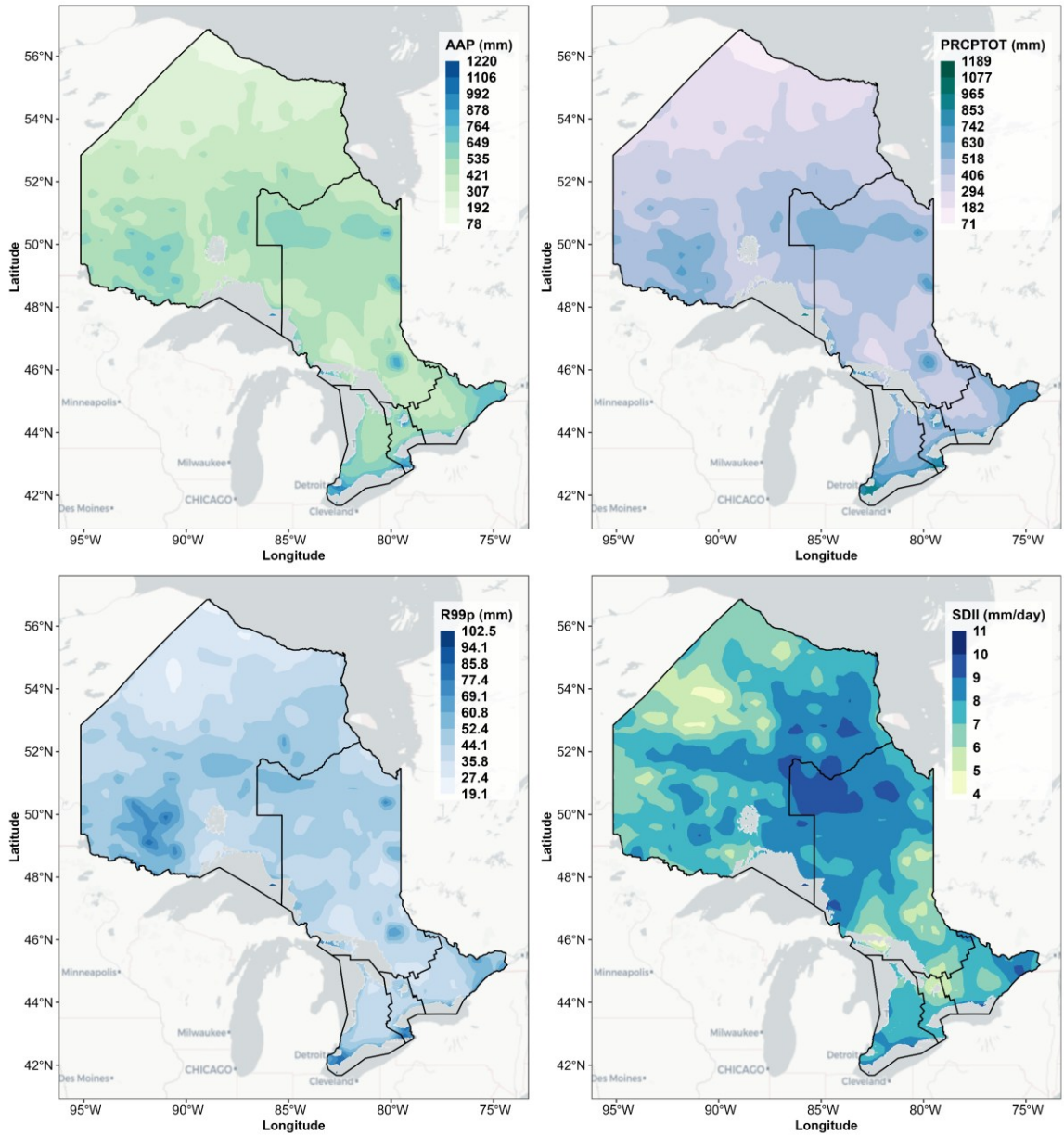

**Figure 5.** Average annual precipitation (AAP) in mm based on the NOAA CMORPH data covering 1998-2023. Lakes and water bodies are excluded from the map. Categorized values of the selected climate indices (a) PRCPTOT, (b) R99p, and (c) SDII.

For the second sub-component of meteorological hazard, projected changes in extreme rainfall across Ontario were first assessed using a comprehensive dataset that included 144 gauged stations and 67 ungauged locations. The primary data source utilized is the IDF_CC Tool (Version 7.5,

June 2024), developed by the University of Western Ontario and the Institute for Catastrophic Loss Reduction (Schardong et al., 2020). This web-based platform incorporates statistically downscaled climate projections from CMIP6 (26 GCMs) ensembles across various emission scenarios (https://www.idf-cc-uwo.ca/). Although Environment and Climate Change Canada (ECCC) provides IDF curves for many locations nationwide (https://climatedata.ca/), its datasets do not include reliable future projections for substantial areas of Northern Ontario. As such, the IDF_CC Tool is deemed more suitable for this analysis. However, the tool exhibits inherent limitations, particularly in ungauged locations, as highlighted by Gaur et al. (Gaur et al., 2020), including sensitivity to the selection of reanalysis datasets, machine learning algorithms, and spatial interpolation methodologies. These factors contribute to uncertainty in the projected IDF estimates and warrant cautious interpretation.

Future rainfall projections are generated for the entire province under two key emission scenarios from the IPCC Sixth Assessment Report: SSP2-4.5, which represents a "middle-of-the-road" development pathway with moderate challenges to mitigation and adaptation, and SSP5-8.5, which reflects a fossil-fuel-intensive, high-emissions trajectory. These Shared Socioeconomic Pathways (SSPs) are standardized scenarios used in CMIP6 climate modeling to explore a range of plausible future socioeconomic and emissions conditions, and to facilitate consistent comparison across climate impact studies. Spatially continuous datasets of both historical and projected extreme rainfall are developed for two storm durations (12-hour and 24-hour), covering mid-century (2041–2070) and late-century (2071–2100) time horizons. Seven standard return periods (2-, 5-, 10-, 20-, 25-, 50-, and 100-year) are incorporated. The resulting data are processed using kriging-based spatial interpolation techniques to create high-resolution spatial maps that illustrate the distribution of future rainfall extremes across Ontario. Percentage changes relative to historical baselines are quantified and visualized. Figure 6 shows the projected percentage changes of 12 h 100-year rainfall under the two emission scenarios for mid and distant future.

For integration into the risk assessment framework, the percentage increase in rainfall relative to historical conditions is used, specifically focusing on projected 12-hour, 100-year rainfall events under medium (SSP2-4.5) and high (SSP5-8.5) emission scenarios for the late-century period (2071–2100) (Figure 6, bottom row). The 12-hour duration is commonly used in infrastructure design, especially for urban and transportation drainage systems, as it captures prolonged storm events that can lead to significant surface runoff and localized flooding. The 100-year return period corresponds to a high-impact, low-probability event, aligning with industry standards for designing resilient infrastructure. It ensures preparedness for extreme but plausible rainfall scenarios, which are increasingly important under changing climate conditions. The choice of the SSP5-8.5 scenario, a high-end emission pathway, is made to account for worst-case projections. It provides a conservative basis for long-term adaptation planning, especially for assets with long service lives such as highways and stormwater systems. As will be detailed further below, rainfall projections are applied in the final step of our risk assessment framework, where they are used to derive both the maximum projected rainfall increase and the risk-based adjusted percentage changes for the return period and duration of interest.

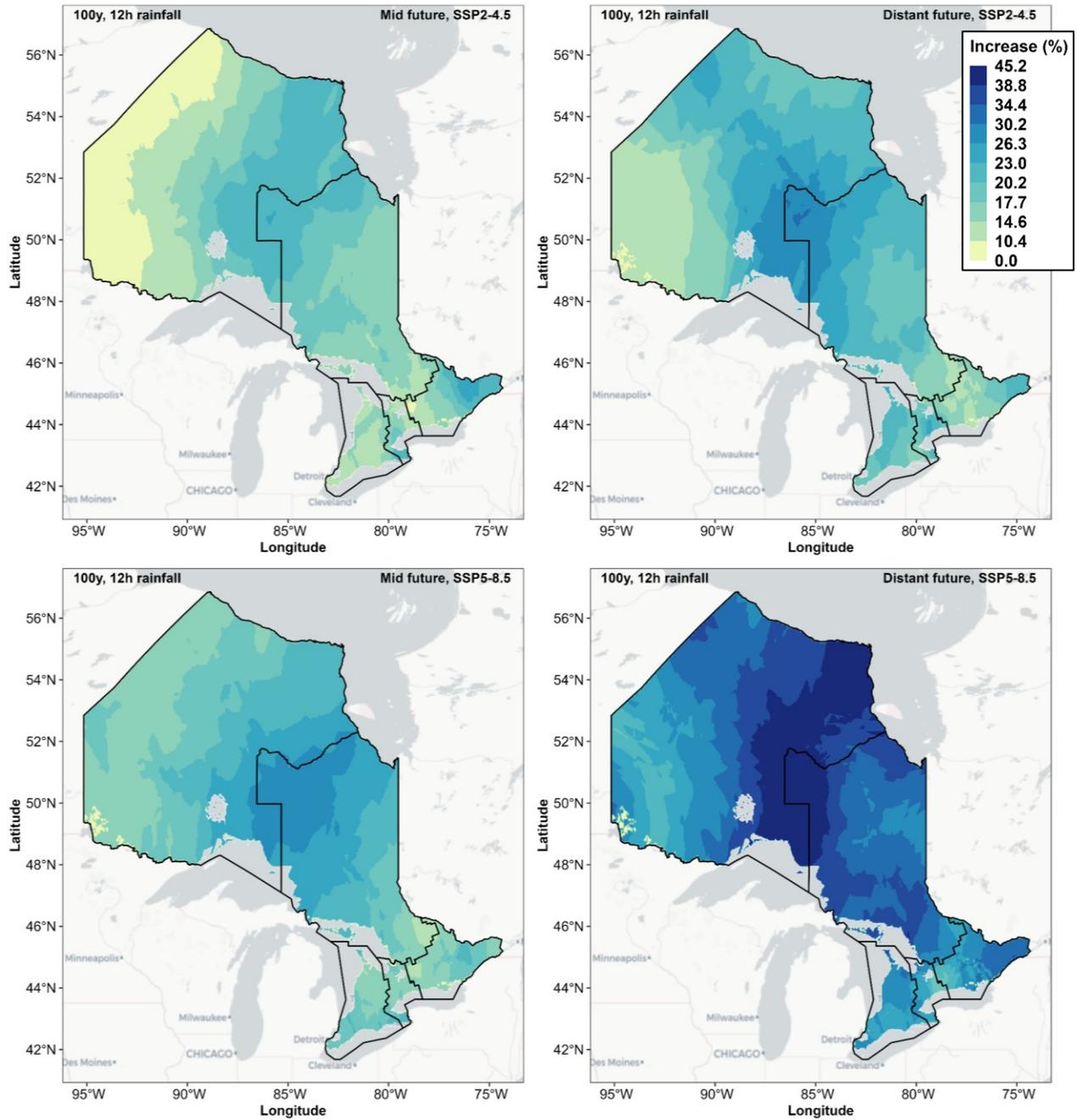

**Figure 6.** Generated maps of projected percentage changes of 12 h 100-year rainfall under the (left column) SSP2-4.5 and (right column) SSP5-8.5 scenarios for (top row) mid future (2041-2070) and (bottom row) distant future (2071-2100).

### 3.1.2. Vulnerability level

While hazards describe the potential for flooding, vulnerability defines the degree to which human, infrastructural, and ecological systems are at risk. Vulnerability is shaped by a combination of exposure, sensitivity, and adaptive capacity. Exposure refers to the presence of people, assets, or ecosystems in flood-prone areas. Sensitivity indicates how susceptible these elements are to flood damage, based on structural integrity, economic resilience, and environmental sensitivity. Adaptive capacity reflects the ability of systems to anticipate, respond to, and recover from flood impacts.

When it comes to highway infrastructure, vulnerability can be defined as the susceptibility of roads, bridges, culverts, and drainage systems to flood-induced disruptions. This vulnerability arises from both physical factors, such as inadequate drainage capacity, aging infrastructure, and poor road materials, and operational factors, including maintenance frequency. Beyond structural concerns, functional vulnerability significantly influences overall risk. It reflects the extent to which disruptions in highway networks impact mobility, emergency response, and economic activities. Major transportation corridors, such as those supporting freight movement or critical supply chains, are more vulnerable because their failure can lead to widespread economic and logistical consequences. Conversely, highways with well-integrated detour routes and redundancy in the transportation network may exhibit lower functional vulnerability despite being physically exposed to flooding.

In this paper, the vulnerability level is categorized into three major aspects, as defined below:

- **Socio-economic vulnerability** accounts for population density and economic resilience in flood-prone areas.

- **Transportation vulnerability** is introduced as a context-specific component that considers traffic volume (or road classification), vehicle type, and the availability of alternate routes, which together influence network resilience.

- **Environmental vulnerability** is characterized by contamination risk, erosion risk, and habitat disruption, reflecting the ecological consequences of extreme flood events.

These risk components provide a basis for assessing the flood susceptibility of the infrastructure and adjacent area which will be impacted at the time of failure. In this paper, transportation vulnerability is measured in terms of the consequence of failure of the drainage infrastructure. This approach implicitly accounts for intrinsic and physical vulnerability, based on the assumption that roads with higher functional importance, such as freeways and arterials, not only contribute more substantially to network connectivity but are also typically associated with more complex, costly, and structurally robust designs.

Flooding can also have severe socio-economic impacts, particularly in densely populated urban centers. Figure S4 displays Ontario's gridded population density classified into five quantile-based levels, derived from the global gridded dataset by NASA's Socioeconomic Data and Applications Center (SEDAC) (CIESIN, 2018). Higher population density increases exposure to flood hazards, as more people rely on transportation networks for daily commutes, emergency evacuation, and economic activity. To capture this dimension of vulnerability, we utilize the Canadian Social Vulnerability Index (SVI), which was specifically developed to assess vulnerability to natural hazards as part of the National Human Settlement Layer (NHSL) (Journeay, 2022). The SVI captures community-level capacity to endure and recover from disaster events, using indicators that reflect underlying characteristics such as housing conditions, household composition, individual independence, and economic resilience. The original SVI dataset (available at https://github.com/OpenDRR/national-human-settlement) is in a 5 km hexagonal grid, from which data for Ontario has been extracted, resampled, and converted to a raster format at 1 km spatial resolution to maintain consistency with other geospatial layers used in this paper. Having classified the SVI into 5 levels, as shown in Figure 7, there is substantial spatial variability in vulnerability levels across the region. The highest vulnerability levels are concentrated in Southern Ontario, particularly in densely populated urban centers such as the Greater Toronto Area (GTA), Hamilton, and London. In contrast, Northern and Western Ontario exhibit lower vulnerability levels, reflecting lower population densities and fewer urbanized areas. Some regions in Eastern Ontario and along major transportation corridors show moderate vulnerability, likely due to a mix of urban-rural transitions and economic disparities.

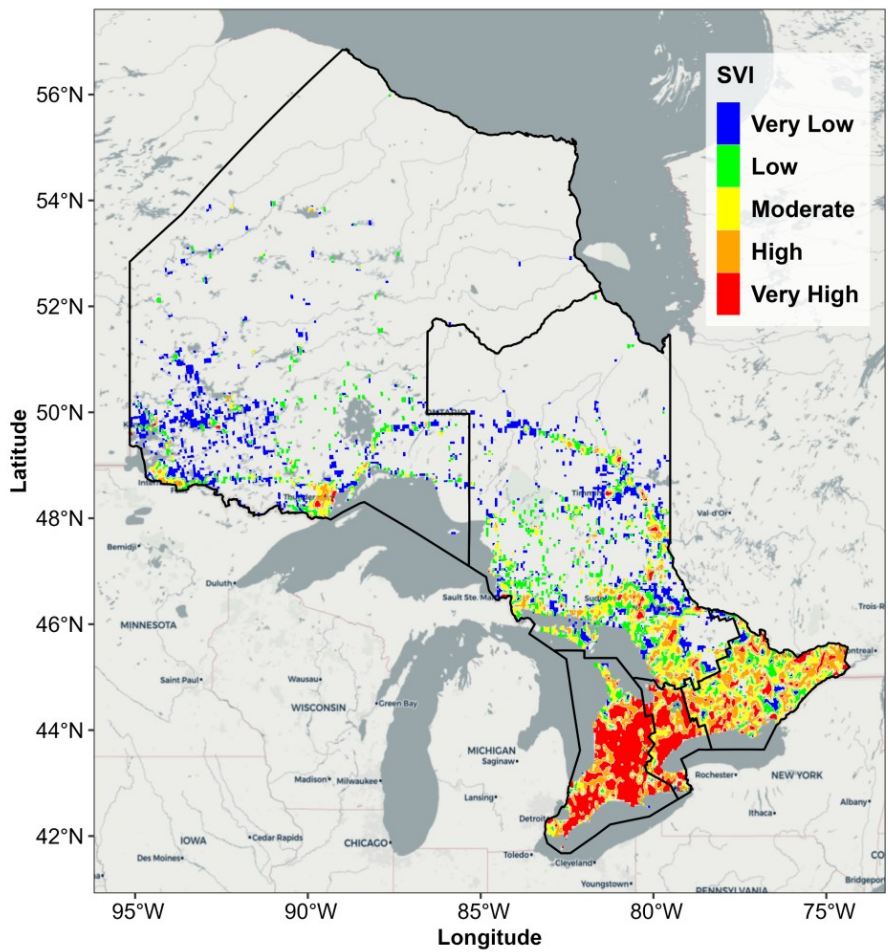

**Figure 7.** Categorized SVI level using the quantile-based approach described in the text.

Transportation vulnerability is defined based on three key factors: road classification, vehicle type, and the availability of alternate routes. These factors are evaluated in the context of potential drainage infrastructure failure, acknowledging that impacts can differ substantially depending on the criticality of the roadway, the types of vehicles relying on it, and the presence of viable detour options. While the general concept of assessing consequences—such as impacts on public safety, traffic delays, flood-related damage, and natural habitat disruption—is aligned with the criteria outlined in the MTO Highway Drainage Design Standards (MTO, 2023a), the specific formulation, structure, and categorization of the condition-level tables are a novel contribution of this study. These tables are designed to help practitioners systematically assess and categorize transportation vulnerability based on the specific characteristics and criticality of the road segment.

Table 2 presents a structured five-level framework for assessing transportation vulnerability across three key factors: road class, vehicle type, and alternate road availability. Road classes reflect their functional significance, with freeways (e.g., 400-series highways) assigned critical vulnerability due to their central role in regional connectivity, followed by arterials (very high vulnerability), collectors (high vulnerability), local roads (medium vulnerability), and seasonal roads (low vulnerability). Vehicle vulnerability ranges from low for passenger cars and light trucks to critical for emergency and agricultural vehicles, with intermediate categories reflecting increasing importance in freight, infrastructure, and logistics. Alternate road vulnerability is based on the availability and quality of substitute routes, progressing from low vulnerability with extensive, high-quality backups to critical vulnerability when no viable alternatives exist.

**Table 2.** Condition-level table for transportation vulnerability components.

| Factor | Name | Description | Vulnerability level |
|---|---|---|---|
| Road Class | Seasonal Roads | Roads maintained seasonally, often in rural or recreational areas | Low (1) |
| | Local Roads | Low-capacity roads for local access and neighborhood connections | Medium (2) |
| | Collectors | Moderate-capacity roads linking arterials and local roads | High (3) |
| | Arterials | High-capacity roads connecting urban areas and major routes | Very high (4) |
| | Freeways | Controlled-access highways, e.g., 400-series highways | Critical (5) |
| Vehicle Type | Light | Passenger cars, motorcycles, SUVs, and light trucks | Low (1) |
| | Medium | Small delivery vehicles and light-duty commercial trucks | Medium (2) |
| | Heavy Vehicles | Transport trucks, buses, and construction equipment | High (3) |
| | Critical Logistics | Fuel transport trucks, medical supply trucks, and essential infrastructure maintenance vehicles | Very high (4) |
| | Special Vehicles | Emergency vehicles, agricultural machinery, etc. | Critical (5) |
| Alternate Road | Extensive Alternative Access | Multiple high-quality, MTO-classified alternative routes available | Low (1) |
| | Viable Backup Routes | One or two viable MTO-classified alternative routes | Medium (2) |
| | Limited Functional Options | A limited number of alternative routes available, but with moderate capacity and acceptable condition | High (3) |
| | Poor Alternative Access | Few alternatives with limited capacity or poor condition | Very high (4) |
| | No Backup Access | No alternative routes; heavy reliance on the primary road | Critical (5) |

Similar to transportation vulnerability, we have identified three primary dimensions of environmental vulnerability relevant to drainage infrastructure: water contamination, erosion and sedimentation, and disruption of natural habitats. The extent of an area's vulnerability to these factors is assessed in the context of drainage infrastructure failure. The potential environmental consequences of such failures dictate the level of vulnerability that must be accounted for in risk assessments. It is worth noting that public safety and disruption of natural habitats are somewhat analogous to factors outlined in the MTO design guidelines for Temporary Drainage Facilities (MTO, 2023a), which assist practitioners in determining the minimum annual exceedance probability for design flows. We integrated these perspectives into the development of a structured condition-level framework for evaluating environmental vulnerability.

Table 3 outlines a five-level classification system for environmental vulnerability for the three components mentioned above (water contamination, erosion and sedimentation, and natural habitat disruption). For water contamination, vulnerability ranges from negligible risk (Class 1) to catastrophic, long-term impacts (Class 5), reflecting increasing pollutant severity and potential ecological and public health consequences. Erosion and sedimentation vulnerability similarly progresses from minimal erosion (Class 1) to catastrophic sedimentation (Class 5), with higher classes indicating escalating disruption to land, water, and ecosystems. Natural habitat disruption follows the same ascending gradient, from localized disturbances (Class 1) to irreversible damage to critical ecosystems (Class 5), indicating rising levels of habitat loss and ecological degradation.

**Table 3.** Condition-level table for environment vulnerability components.

| Factor | Name | Description | Vulnerability level |
|---|---|---|---|
| Water Contamination | Negligible Contamination | Minimal risk of pollutants entering water bodies | Low (1) |
| | Minor Contamination | Some risk of minor pollutants contaminating water | Medium (2) |
| | Moderate Pollutant | Significant risk of pollutants affecting water | High (3) |
| | High-Risk Toxic | Severe risk with toxic or hazardous materials | Very high (4) |
| | Long-Term Contamination Impact | Catastrophic contamination with long-term effects | Critical (5) |
| Erosion and Sedimentation | Minimal Erosion | Minimal erosion with negligible sediment deposition | Low (1) |
| | Manageable Erosion | Moderate erosion with manageable sediment effects | Medium (2) |
| | Significant Erosion | Significant soil erosion affecting land and water | High (3) |
| | Severe Ecosystem Disruption | Severe erosion disrupting ecosystems and agriculture | Very high (4) |
| | Catastrophic Sedimentation | Catastrophic erosion with widespread sedimentation | Critical (5) |
| Natural Habitat Disruption | Localized Disturbance | Minimal disruption to local habitats | Low (1) |
| | Reversible Habitat Impact | Partial habitat alteration with recoverable effects | Medium (2) |
| | Critical Habitat Disruption | Significant disruption to critical habitats | High (3) |
| | Extensive Habitat Loss | Severe and widespread habitat destruction | Very high (4) |
| | Ecosystem Irreversibility | Irreversible damage to key ecosystems | Critical (5) |

## 3.2. Relative importance of risk components

To investigate the influence of risk components in our provincial-wide assessment, where input spatial maps are available, we employed a structured, scenario-based weighting scheme. This approach allows for a repeatable sensitivity analysis, examining how different weight distributions between contributing factors affect the spatial distribution of risk across Ontario's road network. Each scenario represents a distinct configuration of weights assigned to paired components within the hazard or vulnerability domains. For hazard, we evaluated combinations such as terrain versus runoff potential (CN values) under the TxCx scenario group; Historical extreme climate indices versus future rainfall extremes under the XxFx group; and meteorological drivers versus physiographic characteristics under the MxPx group. Similarly, for vulnerability, we considered the relative importance of social vulnerability (SVI) versus road classification under the SxRx scenario group.

For each group, we tested multiple weighting combinations, including equal importance (0.5/0.5) and component-skewed scenarios (0.7/0.3 and 0.3/0.7), to evaluate how risk outputs shift under different assumptions about the driving factors. For example, the S7R3 scenario assigns 70% weight to SVI and 30% to road class, emphasizing the social dimension of vulnerability, while S3R7 flips that balance. This method enables a practical and systematic sensitivity analysis without relying on subjective expert inputs. A complete list of the scenario codes and their associated weightings is summarized in Table 4. The donut plots in Figures 9, 11, and 12 illustrate regional-scale variations in hazard levels across three selected weighting configurations. These visualizations highlight how different weighting choices can alter the proportional distribution of hazard levels. Practitioners can use this insight, together with local context, to select the most appropriate configuration for their needs.

### 3.3. Risk Level

Flood hazard level and vulnerability level are combined to determine the overall risk, which is represented using a Flood Risk Matrix, which visually represents the interaction between hazard ($H$) levels and vulnerability ($V$) levels. $H$ is computed as

$$H = W_{phys} \cdot H_{phys} + W_{met} \cdot H_{met} \quad (5)$$

$W_{phys}$ and $W_{met}$ are the weights assigned to physiographic and meteorologic hazards. $V$ represents the vulnerability level, calculated as

$$V = W_{soc} \cdot V_{soc} + W_{trans} \cdot V_{trans} + W_{env} \cdot V_{env} \qquad (6)$$

where $W_{soc}$, $W_{trans}$, $W_{env}$ represent the weights assigned to social (SVI), transportation, and environmental vulnerability components, respectively. Each hazard component $H_{phys}$, $H_{hydro}$, $V_{soc}$, $V_{trans}$, $V_{env}$ is further broken down into sub-criteria, with individual weights assigned.

To maintain consistency across diverse indicators in the flood risk assessment, all components are spatially evaluated and normalized using a quantile-based classification. Each variable is ranked and assigned to one of five discrete levels based on its percentile distribution: very low (0–20$^{th}$ percentile), low (20$^{th}$–40$^{th}$), moderate (40$^{th}$–60$^{th}$), high (60$^{th}$–80$^{th}$), and very high (80$^{th}$–100$^{th}$). This classification is applied to both hazard and vulnerability indicators.

Table 4. Scenario configuration and weights for hazard and vulnerability components

| Scenario Group | Scenario Code | Component 1 | Component 2 | Weight (Comp 1) | Weight (Comp 2) |
|---|---|---|---|---|---|
| Hazard – Terrain vs Runoff | T3C7 | Terrain features | Curve Number (CN) | 0.3 | 0.7 |
| | T5C5 | | | 0.5 | 0.5 |
| | T7C3 | | | 0.7 | 0.3 |
| Hazard – Climdex vs Future Rainfall | X3F7 | Climdex (Extreme Indices) | Future Rainfall Projections | 0.3 | 0.7 |
| | X5F5 | | | 0.5 | 0.5 |
| | X7F3 | | | 0.7 | 0.3 |
| Hazard – Meteorology vs Physiography | M3P7 | Meteorological Factors | Physiographic Factors | 0.3 | 0.7 |
| | M5P5 | | | 0.5 | 0.5 |
| | M7P3 | | | 0.7 | 0.3 |
| Vulnerability – SVI vs Road Class | S3R7 | Social Vulnerability Index (SVI) | Road Class | 0.3 | 0.7 |
| | S5R5 | | | 0.5 | 0.5 |
| | S7R3 | | | 0.7 | 0.3 |

Considering the risk matrix shown in Figure 8, regions with low hazard and vulnerability are categorized within the green zone, signifying minimal flood risk. Areas exhibiting moderate hazard and vulnerability fall into the yellow-orange zone, indicating the need for proactive risk management strategies. Conversely, regions with very high hazard and vulnerability are classified in the red zone, necessitating a more conservative design storm.

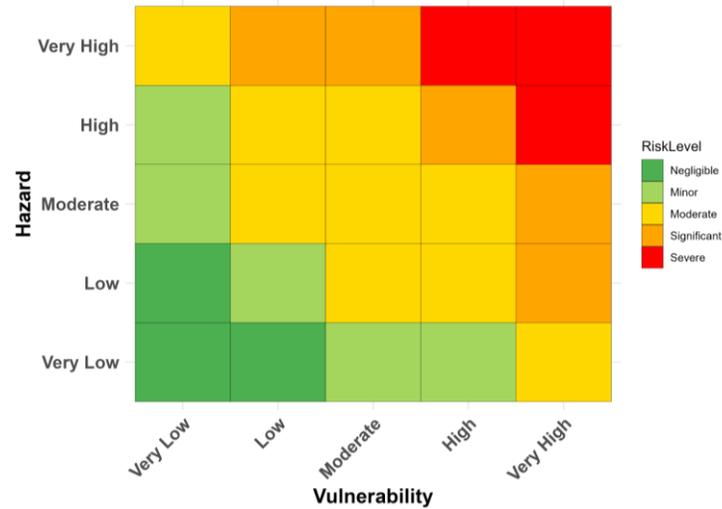

**Figure 8.** Flood risk matrix.

**3.4. Risk-based future design storm**

We propose a linear adjustment approach, whereby the projected percentage increase in future rainfall is proportionally distributed across the five defined risk levels. Let $IDF\%_{mid}$ and $IDF\%_{far}$ denote the projected future rainfall increase (%) for the mid- and far-future periods, respectively, relative to the baseline IDF intensity ($IDF_{Baseline}$), derived from the multimodel ensemble median of a high-emission scenario. Given the non-linear pattern of projected future rainfall changes, $P = max(IDF\%_{mid}, IDF\%_{far})$ represents a reasonable upper bound for the projected increase in rainfall over the entire design lifespan. The rainfall adjustment for each risk level is calculated as shown in Table 5.

**Table 5.** Risk-based adjustment of the future rainfall. $P\,[\max(IDF\%\_mid, IDF\%\_far)]$ is the reasonable upper bound for the projected increase in rainfall over the entire design lifespan.

| Level | Risk class     | Equation                            |
|-------|----------------|-------------------------------------|
| 1     | Very Low Risk  | $P \times 10\% + IDF_{Baseline}$    |
| 2     | Low Risk       | $P \times 25\% + IDF_{Baseline}$    |
| 3     | Medium Risk    | $P \times 50\% + IDF_{Baseline}$    |
| 4     | High Risk      | $P \times 75\% + IDF_{Baseline}$    |
| 5     | Very High Risk | $P + IDF_{Baseline}$                |

This linear adjustment is intentionally conservative. While depth–damage relationships in flood hazard studies often follow nonlinear or exponential patterns, applying such curvature here would lead to steep reductions in the rainfall adjustment for low-risk areas, which could result in under-preparation. By contrast, the proportional scaling provides a gradual reduction from the maximum projected increase while preserving meaningful adjustments for lower risk classes. Moreover, this approach maintains a minimum threshold by enforcing baseline IDF values in cases where projected changes are negative or negligible which prevents underestimation of storm severity in future planning. For the highest risk category, the designer applies the maximum projected IDF values from the SSP5-8.5 multi-model median across both future time horizons, i.e., infrastructure is designed for the most severe climate scenario. This risk-based adjustment framework provides a pathway for climate-informed infrastructure design, while balancing engineering feasibility and cost-effectiveness.

## 4. Province-wide flood hazard potential

Figure 9 presents the spatial distribution of physiographic hazard levels across Ontario, derived from the integration of terrain characteristics and CN maps under three weighting configurations: T7C3, T5C5, and T3C7. The classification ranges from level 1 (very low hazard) to level 5 (very high hazard). Overall, the hazard level distribution is dominated by moderate to high classes (levels 3 and 4), with notable regional differences. In the Northwestern and Northeastern regions, the T3C7 scenario (where CN has higher influence) results in a greater proportion of high-hazard areas (e.g., 39% in Northwestern, 31% in Northeastern). Conversely, increasing the influence of terrain (as in T3C7) shifts the distribution slightly toward more moderate hazard levels in these regions. The Eastern region shows a more balanced hazard profile, particularly under the T5C5 and T3C7 scenarios, where a significant part of the region falls into moderate (level 3) categories, with only a limited extent falling into the very low hazard category under T3C7. In the Central region, a more diverse hazard distribution is observed. While the T7C3 and T5C5 configurations show a dominance of moderate hazard levels (59–64%), the T3C7 scenario reveals a larger share of segments in the high and very high categories (22% combined), indicating increased sensitivity to CN weighting in this area. In the Western region, hazard levels are consistently moderate across all scenarios, with minimal representation in the very high category. These findings highlight the regional variability of flood hazard across Ontario and the importance of integrating terrain and runoff (CN) characteristics. They also reveal how hazard classification is sensitive to weighting schemes, reinforcing the need for context-specific approaches in flood risk assessment.

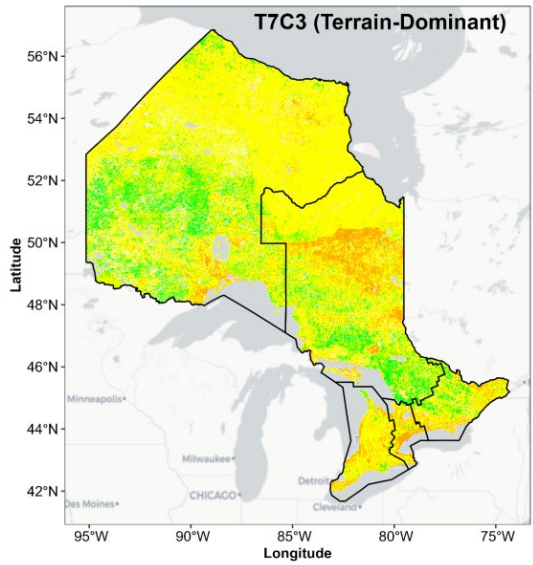
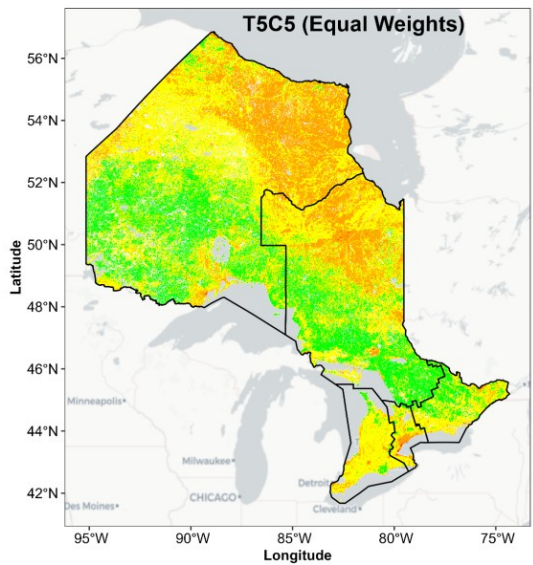
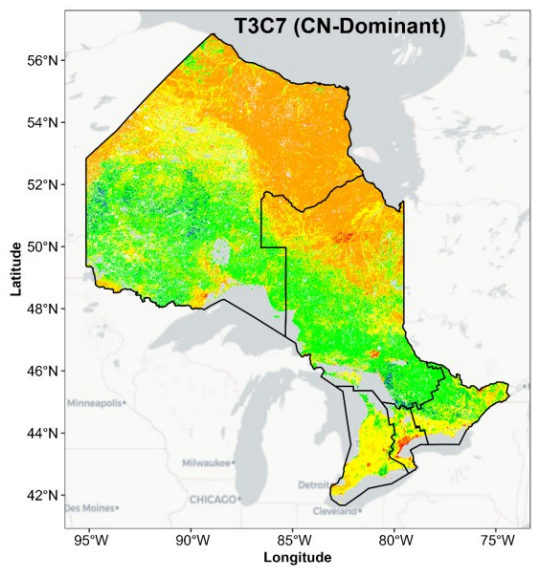
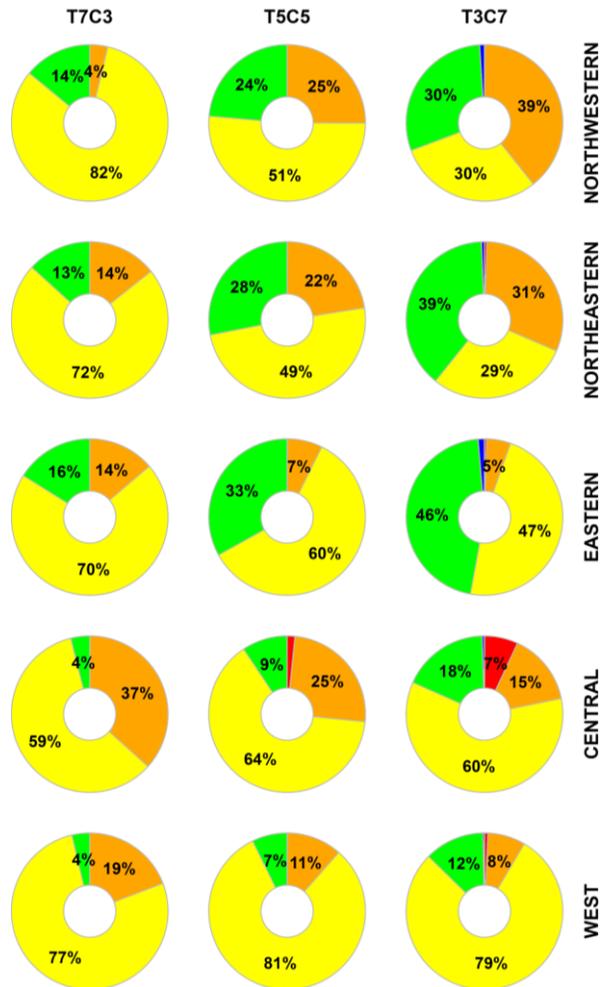

**Figure 9.** Physiographic hazard level resulting from the combination of terrain characteristics and CN maps under three weighting configurations: T7C3, T5C5, and T3C7. Donut charts on the right summarize the proportional distribution of each category (Very Low to Very High) within five major sub-regions of Ontario (Northwestern, Northeastern, Eastern, Central, and West) for each weighting scenario.

Figure 10 presents a spatial distribution of the composite indicator resulting from the combination of three climate extreme indices (which incorporates PRCPTOT, R99, and SDII) and the upper bound of projected rainfall change for the mid- and long-term future horizons. The spatial pattern suggests that the northernmost areas are predominantly characterized by level 1, indicating minimal climate extremes (Figure 10a). In contrast, central and southern regions exhibit higher Climdex levels, with extensive areas of levels 3, 4 and 5, signifying moderate to high climate extremes. The most extreme conditions (level 5) are concentrated in localized patches, particularly in the south and southeast, as well as some isolated areas in the central region. This distribution suggests that climate extremes are more pronounced in the southern regions. The upper bound of the projected rainfall changes (Figure 10a) shows a pattern similar to that of the far-future horizon (see Figure 6). However, in some localized areas, the mid-future period exhibits slightly higher positive change percentages than the far future. This outcome confirms the non-linear nature of projected precipitation variability over time, as previously noted in the literature (Singh et al., 2022).

Figure 11 illustrates the regional distribution of meteorological hazard levels across Ontario under three weighting scenarios. These scenarios represent varying emphasis between Climdex-based historical climate extremes and projected changes in design rainfall intensity, transitioning from past observations (X7F3) to equal weighting (X5F5) and future rain-dominant (X3F7). The Northeastern and Northwestern regions display more distributed patterns. Under X3F7, hazard levels tend to rise, especially in the Northwestern. Within the Eastern region under X3F7, the very low and low categories dominate at 63%, with a notable shift from lower classes. This reflects moderate sensitivity to future projections in the East. In the Western region, increasing the weight of future rainfall projections results in a pronounced shift from moderate to lower hazard levels. The moderate hazard class drops from 57% in X7F3 to just 11% in X3F7, while the share of low hazard categories rises substantially. A similar trend emerges in the Central region, where low and very low hazard levels collectively reach 95% under X3F7. These results suggest that regions currently experiencing more frequent historical extremes may not face proportionally higher future risks under projected rainfall scenarios, which highlights the importance of integrating both historical and forward-looking data to avoid over- or underestimation in risk-informed design strategies.

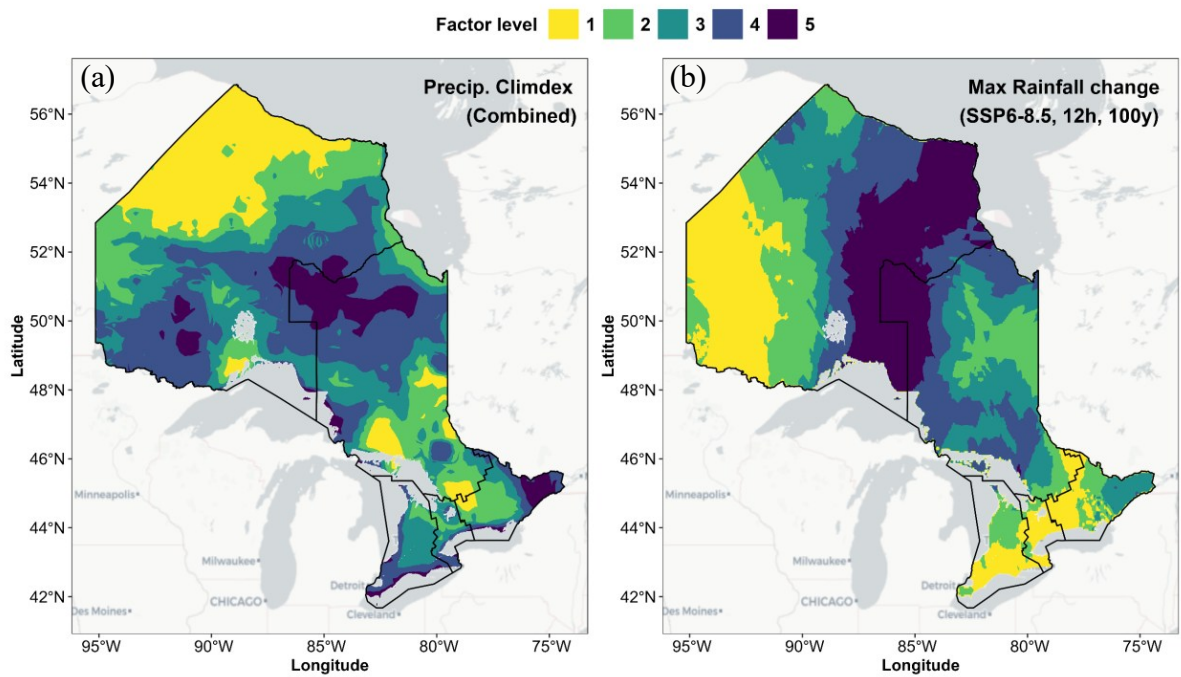

**Figure 10.** Combined precipitation Climdex (a) and the upper bound of projected rainfall change across mid and distant future (b).

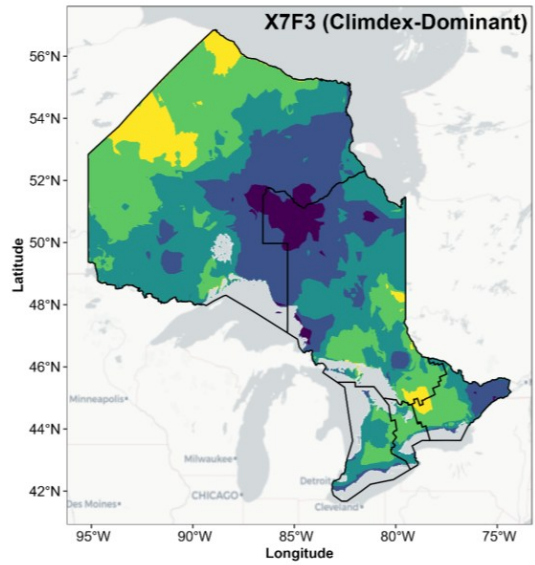
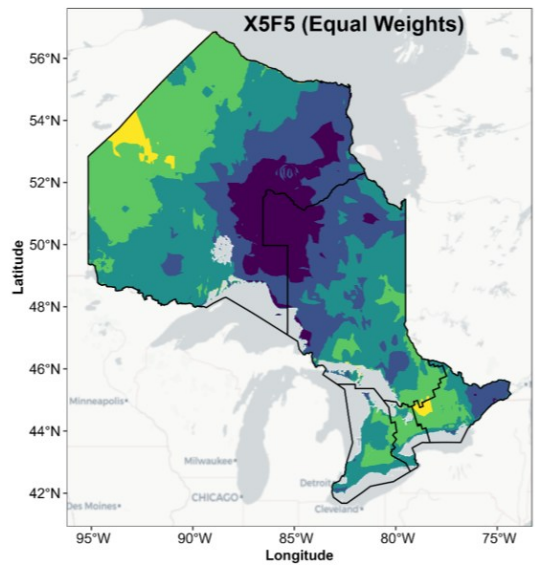
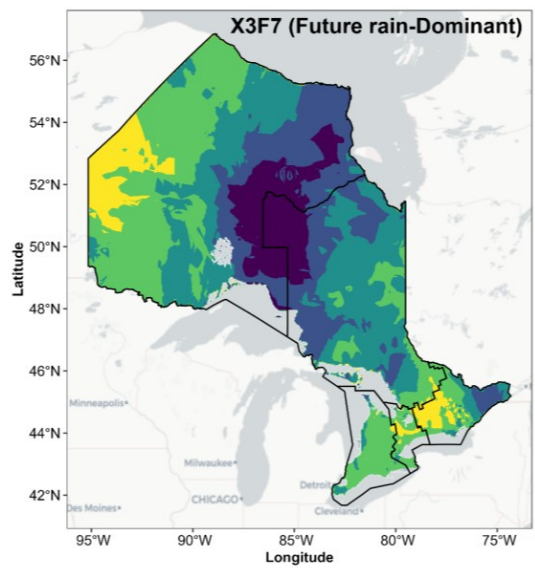
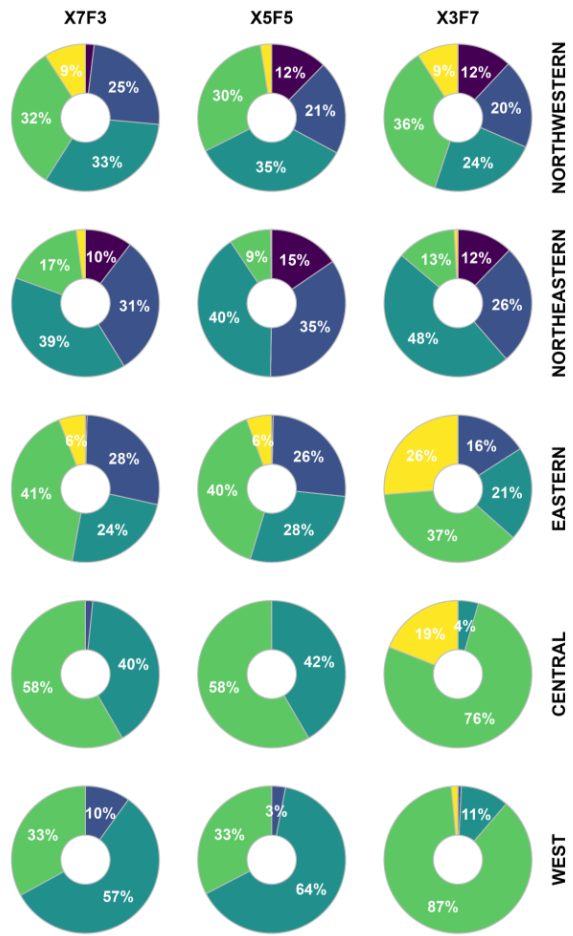

**Figure 11.** Meteorological hazard level derived from the combination of precipitation Climdex, and projected rainfall increase under three weighting configurations: X7F3, X5F5, and X3F7. Donut charts on the right summarize the proportional distribution of each category (Very Low to Very High) within five major sub-regions of Ontario (Northwestern, Northeastern, Eastern, Central, and West) for each weighting scenario.

Figure 12 illustrates the spatial distribution of hazard levels across Ontario under three weighting scenarios (M7P3, M5P5, M3P7), representing different emphases on historical meteorological conditions (M) versus physiographic factors (P). Across all regions, a clear shift toward more moderate and higher hazard levels occurs as the weight of physiography increases. The Northeastern and Northwestern regions, which already show elevated hazard levels under the M7P3 scenario, exhibit a further escalation under M3P7, with the combined proportion of moderate to high hazard exceeding 80%. In the Eastern region, the combined moderate and high hazard levels also grow significantly—from 56% in M7P3 to 70% in M3P7, indicating a clear shift away from lower hazard classifications. The Central region exhibits a similar pattern, with moderate to high hazard reaching 91% under M3P7, compared to only ~57% in M7P3. In the Western region, increasing the weight assigned to physiographic hazard raises the proportion of moderate hazard areas to 94%, substantially reducing the extent of low hazard zones. This indicates that, in this region, physiographic conditions, such as terrain and runoff potential, play a dominant role in shaping overall flood hazard, potentially overshadowing meteorological influences. This suggests that in terrain-sensitive areas, infrastructure design should place stronger emphasis on physiographic vulnerability, especially where intense rainfall may exacerbate runoff responses. Ignoring these features could lead to underestimation of hazard, particularly in regions like Central and Western Ontario, where physiographic factors substantially elevate risk when prioritized.

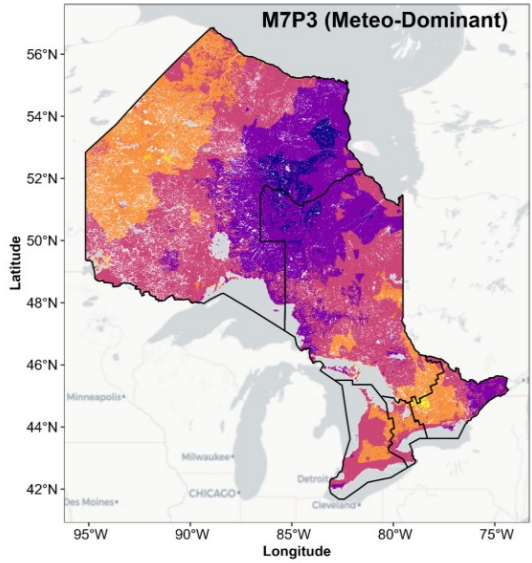
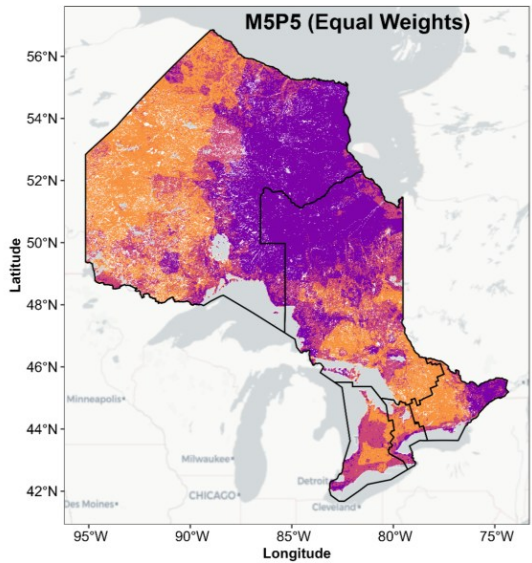
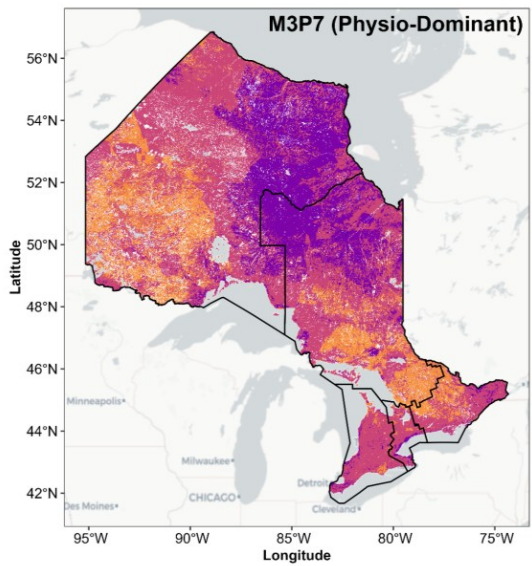
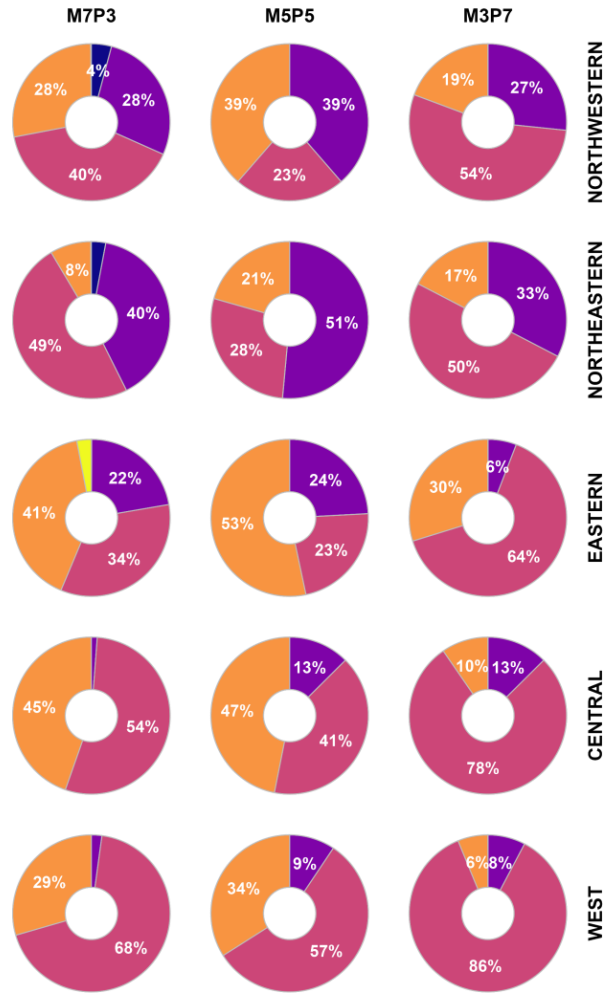

**Figure 12.** Combined hazard level based on physiographic and meteorologic components under three weighting configurations: M7P3, M5P5, and M3P7. Donut charts on the right summarize the proportional distribution of each category (Very Low to Very High) within five major sub-regions of Ontario (Northwestern, Northeastern, Eastern, Central, and West) for each weighting scenario.

To evaluate the proposed hazard mapping framework, Figure 13 shows the spatial agreement between the hazard map developed using equal weighting and the NRCan Flood Susceptibility Index (FSI) (McGrath and Gohl, 2023). Additional comparisons for alternative weighting scenarios are provided in Supplementary Figure S5. NRCan FSI map is primarily informed by physiographic variables such as digital elevation, land cover, and surficial geology, with historical climate data included as one of several inputs. As outlined in the dataset's documentation, the index was developed using an ensemble machine learning model trained on historical flood events and multiple spatial predictors. The comparison is based on a tolerance of ±1 class and involves various combinations of the components contributing to the final hazard map. Among all approaches, the physiographic hazard map—which integrates CN and terrain features—shows the highest agreement (0.682). This indicates that physiographic indicators effectively represent the underlying flood susceptibility captured by the NRCan map. The CN-based map (0.622) performs better than the terrain-only map (0.575), emphasizing the importance of land surface hydrological properties in flood characterization.

The Climdex-based hazard map shows the lowest alignment (0.486), indicating that climate indices alone may not fully account for the spatial patterns of flood susceptibility. This may also reflect the fact that the NRCan index primarily emphasizes physiographic and land surface characteristics, with limited weighting given to climate-based drivers such as extreme precipitation (McGrath and Gohl, 2023). When combined with physiographic factors (Physiographic + Climdex), the initial agreement decreases to 0.571, implying that the addition of climate indices does not necessarily enhance consistency with the national product. This panel is included to isolate the influence of climate extremes without introducing projected rainfall data, which is not part of the NRCan map. The final proposed hazard map (Physiographic + Meteorologic) shows a slightly higher agreement (0.584) than the Physiographic + Climdex case, but this increase should not be directly attributed to the rainfall projections, as the national benchmark does not account for future conditions. Instead, the observed variations highlight the complexity of integrating dynamic and static indicators in flood hazard mapping.

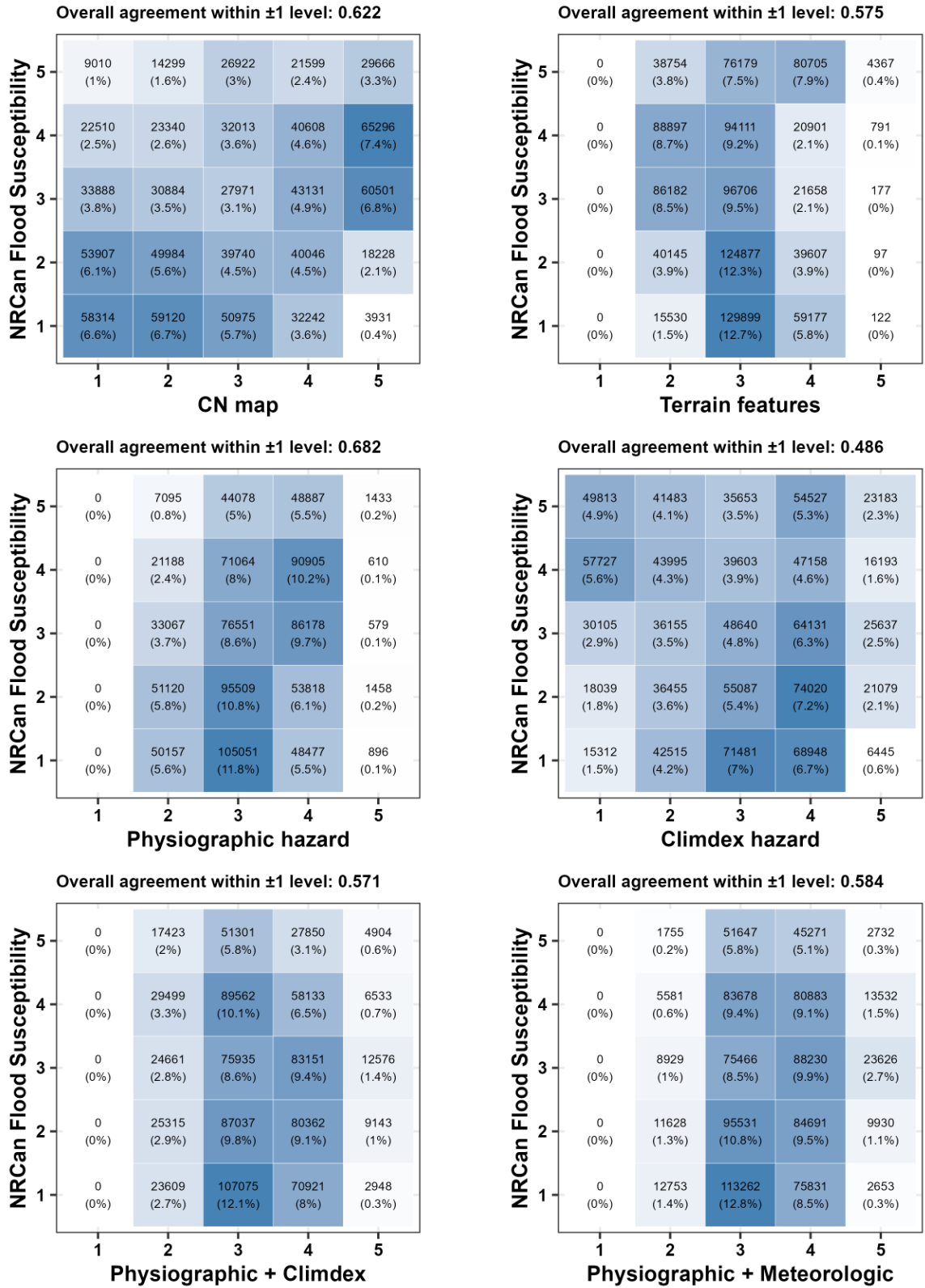

**Figure 13.** Confusion matrix comparing flood hazard levels in this study with the National Flood Susceptibility provided by NRCan. The top (bottom) number in each cell represents

the number of grid points (percentage of total grid points) that fall into the corresponding classes for both datasets. Higher values along the diagonal indicate stronger agreement. The overall classification accuracy is shown above each panel.

## 5. Flood risk in road networks

Building on the hazard assessment, evaluating vulnerability provides the foundation for calculating flood risk across transportation infrastructure using the risk matrix. For the province-wide analysis, we incorporated the SVI and road classification to represent spatial vulnerability. Other vulnerability components are excluded at this scale, as they apply more effectively to site-specific evaluations. With the vulnerability map established, the hazard layer—based on equal weighting of physiographic and meteorological factors—is introduced into the risk matrix to calculate the spatial distribution of risk across road segments. Figure 14 illustrates the resulting vulnerability and risk levels under the equal-weight scenario (S5R5). The results for alternative weighting schemes (S7R3 and S3R7) are provided in Supplementary Figures S6 and S7, with corresponding donut plots illustrating the regional distribution of vulnerability and risk shown in Figures S8 and S9. To further highlight the additional context captured by the SVI compared to using only population density, a comparison of vulnerability maps for road class combined with SVI versus road class combined with population density are also provided in the supplementary file (see Figure S10). As observed, the variability of vulnerability levels is notably lower when population density alone is used, reflecting its narrower socioeconomic scope compared to the SVI, which integrates multiple indicators and provides a more comprehensive assessment of vulnerability.

High vulnerability (levels 4–5) tends to concentrate in Southern Ontario, particularly along dense urban and transportation corridors. Central and Northern Ontario exhibit lower vulnerability levels (1–2), though localized hotspots emerge near key settlements and road segments. The spatial pattern of risk broadly mirrors that of vulnerability, but notable deviations emphasize the role of hazard context in shaping infrastructure risk. Areas with moderate vulnerability may exhibit elevated risk if exposed to substantial flood hazards. For instance, parts of northeastern Ontario—including regions near Sudbury and Timmins—display moderate vulnerability but rank higher in risk due to elevated hazard potential. These distinctions highlight the importance of integrating both hazard and vulnerability in infrastructure prioritization.

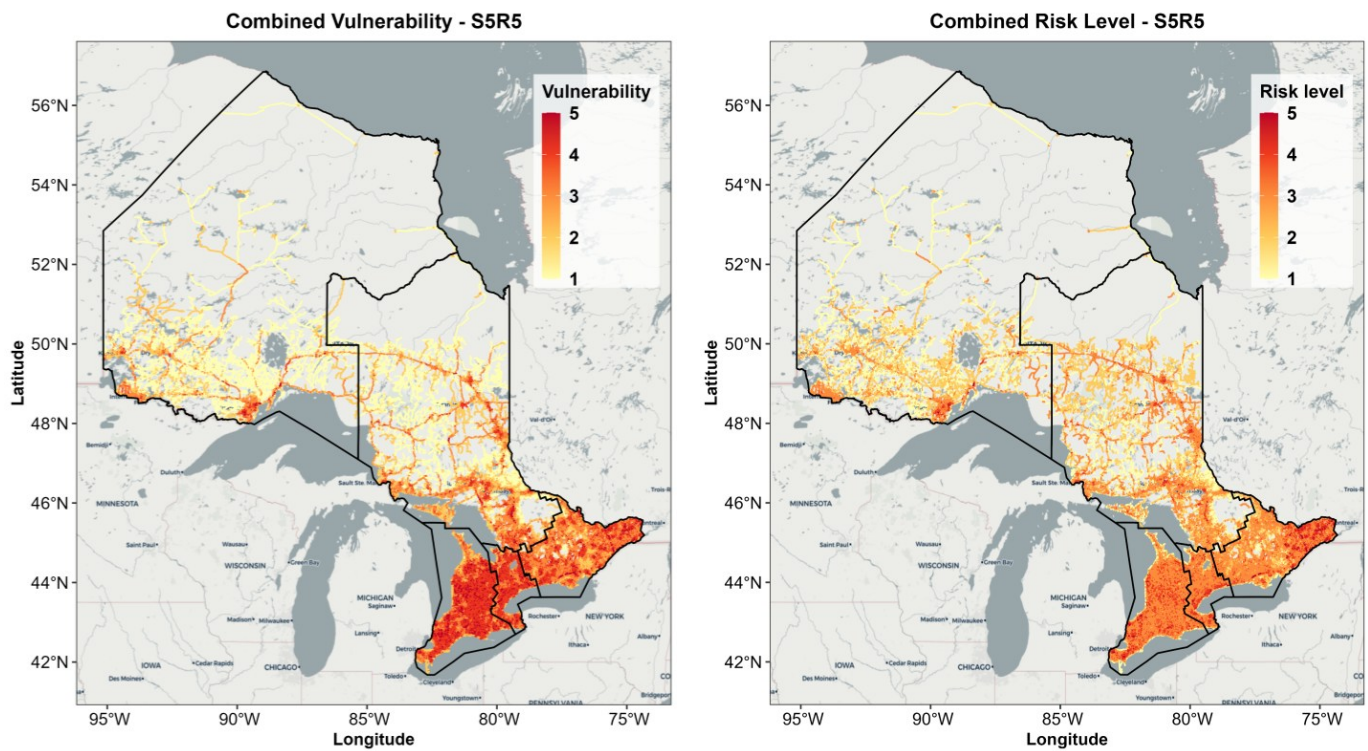

**Figure 14.** Spatial distribution of combined vulnerability and overall risk level across Ontario's road network under scenario S5R5. The left panel displays the composite vulnerability score integrating socio-economic and transportation factors, specifically the Social Vulnerability Index (SVI) and road class, weighted equally. The right panel shows the resulting risk levels based on both hazard intensity and vulnerability scores using a risk matrix. Higher scores highlight road segments that may require greater attention in climate-resilient infrastructure planning.

## 6. Risk-based adjustment maps

Once the risk level is estimated, it becomes possible to apply a risk-based adjustment map to account for the potential changes in future precipitation patterns. These adjustment maps are derived from the upper bound of projections under high-emission climate change scenarios over two mid and distant time horizons. Figure 15 compares the spatial patterns of maximum projected increases in 12-hour 100-year rainfall (left) with the corresponding risk-adjusted rainfall changes (right) under scenario S5R5. The left panel highlights regions in northern and central Ontario experiencing the most intense projected increases, where changes exceed 40% in some areas. However, these projections are largely driven by climate model outputs and do not account for regional vulnerability or exposure. In contrast, the risk-based adjustment map (right) redistributes the projected changes based on localized risk scores. For instance, high-risk areas in Ottawa Valley (within Eastern region) which were less prominent in the raw projections, receive amplified adjustments. Conversely, areas with large projected increases but lower risk, particularly in remote northern regions, experience downward adjustments. For comparison under different weighting schemes, Figure S11 presents the same analysis using scenarios S7R3 and S3R7, which assign different weights to SVI and road class vulnerabilities.

To have better insights into the results, the bottom panel in Figure 15 illustrates the distribution of road segments across MTO regions subjected to varying levels of adjustment, categorized into five percentage brackets (20%, 40%, 60%, 80%, and 100%). Regions such as Central, Eastern, and West show a strong concentration of higher adjustment levels, with over 80% of their segments adjusted by 60% or more. In contrast, the Northwestern region exhibits the lightest adjustment profile, with nearly 60% of road segments falling within the 20–40% adjustment range and less than 9% exceeding 60%, i.e., a substantially lower need for aggressive intervention. Figure S12 illustrates these results under two weighting scenarios, S7R3 and S3R7, representing different emphasis on SVI and road class vulnerabilities, respectively. The results suggest that uniform province-wide standards may lead to inefficient resource allocation and over- or under-adaptation in some areas. The variation in adjustment intensity, driven by spatially resolved risk estimates, underlines the value of risk-informed prioritization frameworks. Rather than treating climate projections as uniform inputs, the use of a risk-adjusted approach enables targeted investment in areas where both exposure and vulnerability converge, maximizing the cost-effectiveness and resilience of future infrastructure upgrades.

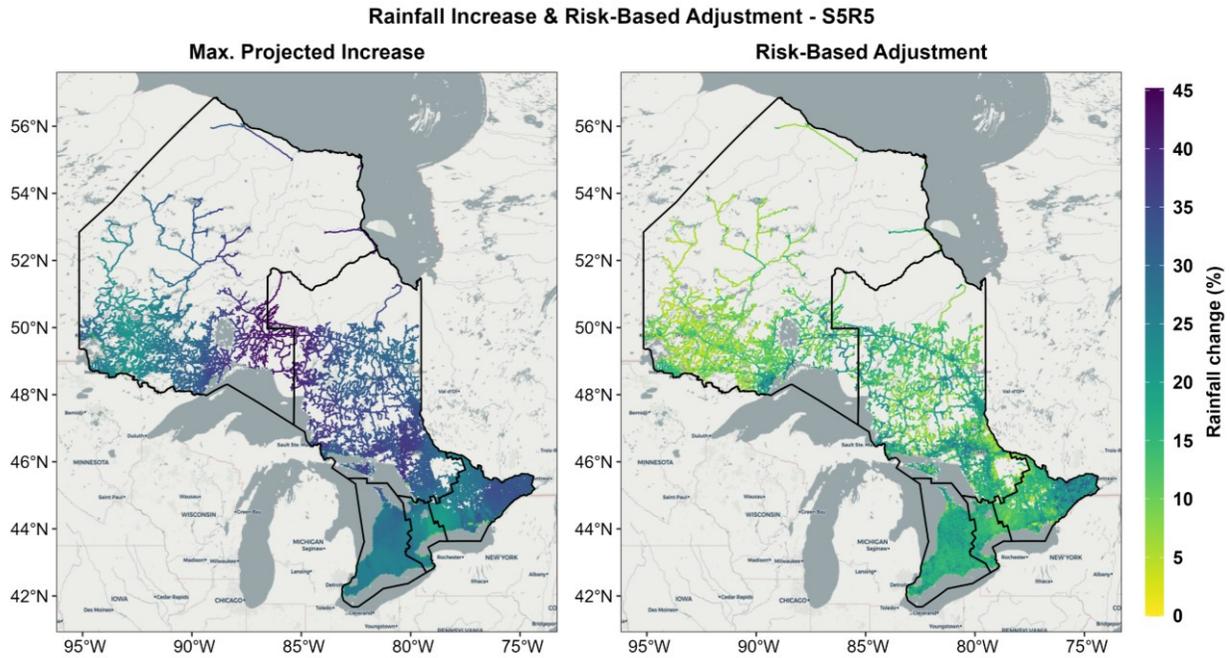

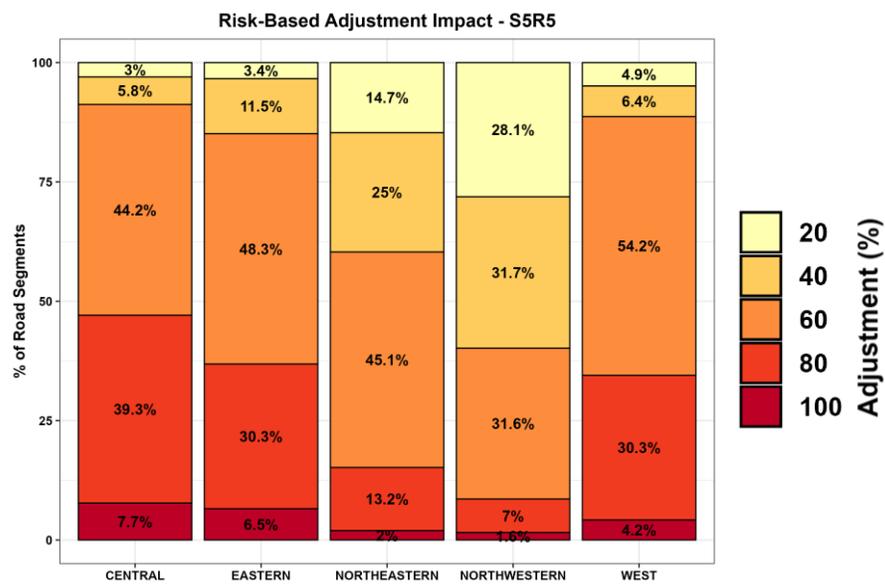

**Figure 15.** Spatial and regional evaluation of future rainfall changes and the impact of risk-based adjustment under scenario S5R5 (equal weights for SVI and Road class vulnerabilities). Top panels illustrate the maximum projected 12-hour 100-year rainfall increase (left) and the adjusted rainfall values based on localized risk scores (right). The bottom panel shows the proportion of road segments within each MTO region subjected to different levels of adjustment.

## 7. Site-specific Implementation

To further demonstrate the proposed methodology, it is applied to three cases of drainage infrastructure to estimate both the risk level and the corresponding adjusted rainfall increase, leveraging available reports on drainage infrastructure across Ontario. The selected locations are within three of five MTO administrative regions. The site-specific assessment approach can be applied to any location, whether for planning new road segments or for rehabilitating existing infrastructure.

The adjusted rainfall increase is applied under the high-emission scenario (SSP5-8.5), specifically for a 12-hour duration and a 100-year return period. The hazard level is directly extracted from the hazard map. SVI—used as a proxy for socio-economic vulnerability—is also determined through its corresponding map. Therefore, latitude and longitude coordinates are the sole data input requirement for the infrastructure of interest (for both hazard and vulnerability). Other contributing factors, which require qualitative assessment, are evaluated based on designer's input and the predefined condition-level tables. We reviewed the available reports and considered the geographic context to determine the vulnerability levels. In cases where reports lacked sufficient information, a medium-level assumption is adopted from the condition-level tables to ensure consistency in risk estimation.

Projected changes in 12-hour rainfall extremes vary notably across the three studied locations While increases are generally observed with return period, future time horizon, and emission scenario, some exceptions occur—particularly at intermediate return periods (Figure 16). For additional insight, corresponding results for 24-hour rainfall extremes are also provided in Figure 17. At C1-West under the SSP2-4.5 scenario, mid-future changes range from approximately 9% (2-year) to 14% (100-year), increasing to 14–19% in the far-future. Under the high-emission SSP5-8.5 scenario, the projected increase reaches up to 27% for the 100-year return period by the far-future. The general trend at this location shows an increase in rainfall change percentages with progression from the mid- to far-future period and from the moderate to the high emission scenario. However, the percentage increase stabilizes for return periods of 20 years and longer.

In both the second (C2-Central) and third (C3-NEast) locations, the 100-year return period consistently exhibits the highest projected rainfall change percentage, particularly under the far-future and high-emission (SSP5-8.5) scenario, although the difference compared to the 50-year return period is minimal. In C2-Central, the distinction between the two emission scenarios (SSP2-4.5 and SSP5-8.5) becomes less pronounced at higher return periods, suggesting a convergence in projected impacts under extreme events. In contrast, the third location (C3-NEast) displays the most substantial divergence between mid- and far-future projections. In this location, the rainfall change percentage for the mid-future under SSP5-8.5 closely aligns with the far-future projection under SSP2-4.5. Across all locations, the SSP5-8.5 scenario in the far-future results in the most pronounced increases, especially at higher return periods. A similar pattern is observed for the 24-hour rainfall duration (refer to Figure 17). An exception is noted in the C1-West location, where the mid-future projections indicate a greater increase in 100-yr rainfall compared to the far-future under the same emission scenario.

Table 6 summarizes the results of the practical implementation across the three drainage locations—West, Central, and Northeastern Ontario. All three locations are assigned a moderate overall risk level (Level 3), despite differences in underlying components. The West region exhibited a higher hazard level (4) but relatively lower socioeconomic and environmental vulnerability, whereas the Central region had the highest SVI level (5), indicating greater socioeconomic sensitivity, despite a lower hazard score. The Northeastern location displayed more balanced vulnerability values (3.5 for both transportation and environment), with the highest projected rainfall increase under SSP5-8.5 (32%). This resulted in the greatest risk-based adjustment (19.2%) compared to the other sites. The results indicate that, in these instances, the overall risk level does not justify applying the maximum projected rainfall increase.

This difference significantly influences hydrologic design due to its direct impact on peak discharge estimates. According to the "rational method" (Thompson, 2006), peak discharge ($Q$) is directly proportional to rainfall intensity ($I$). Therefore, a reduction in rainfall intensity leads to a corresponding decrease in design discharge. For instance, reducing the rainfall adjustment from 32% (maximum projection) to 19.2% (risk-based) represents a 40% reduction in the magnitude of the adjustment. This translates to an approximate 10% reduction in peak discharge, based on the proportional relationship defined by the rational method. Such a reduction has tangible implications for culvert design. It may allow for the selection of a smaller culvert diameter, which not only lowers material and excavation costs but also reduces installation complexity, land acquisition requirements, and long-term maintenance demands. These findings highlight the importance of a risk-based approach rather than a blanket application of worst-case projections, which allows for more targeted adaptation strategies in drainage infrastructure planning.

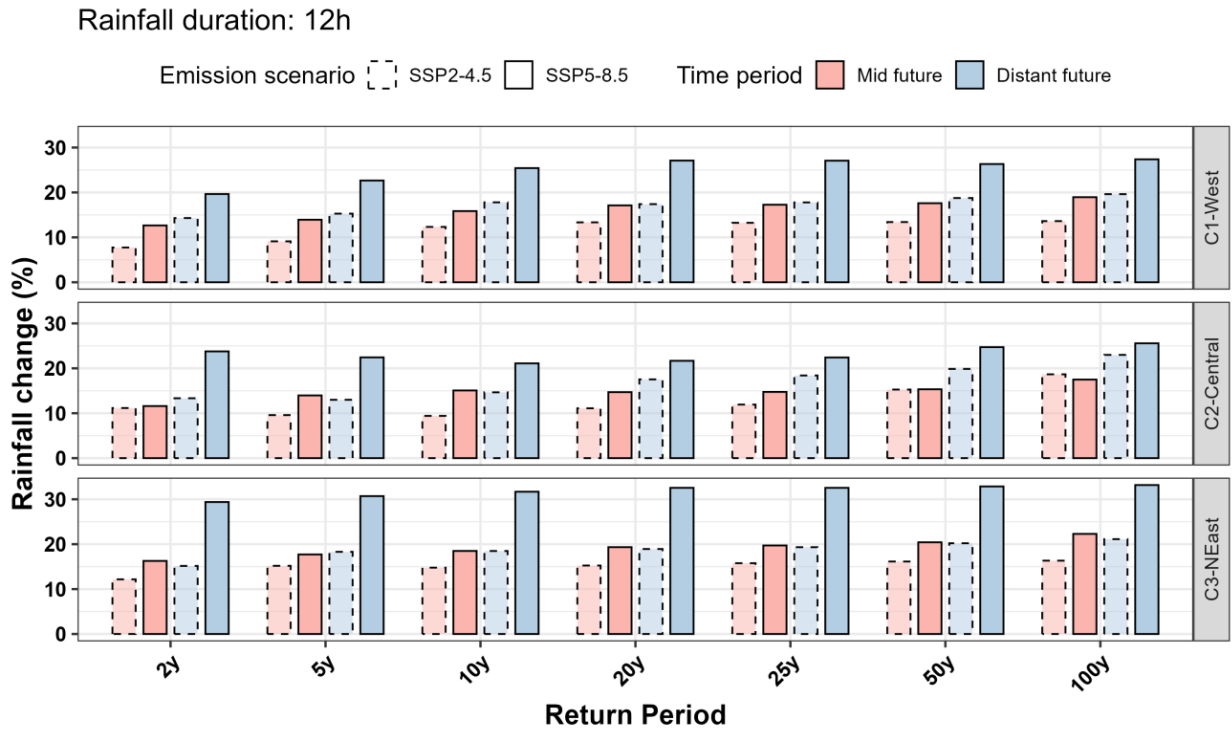

**Figure 16.** Projected changes (%) in 12-hour duration rainfall intensity across seven return periods (2y to 100y) for three selected locations (C1-West, C2-Central, C3-NEast), categorized by emission scenarios (SSP2-4.5, SSP5-8.5) and time periods (Mid-future: 2041–2070, Far-future: 2071–2100). Colored bars represent time periods (red for mid-future, blue for far-future), while bar border styles distinguish emission scenarios (dashed for SSP2-4.5, solid for SSP5-8.5). Rainfall change is expressed as a percentage relative to the historical baseline.

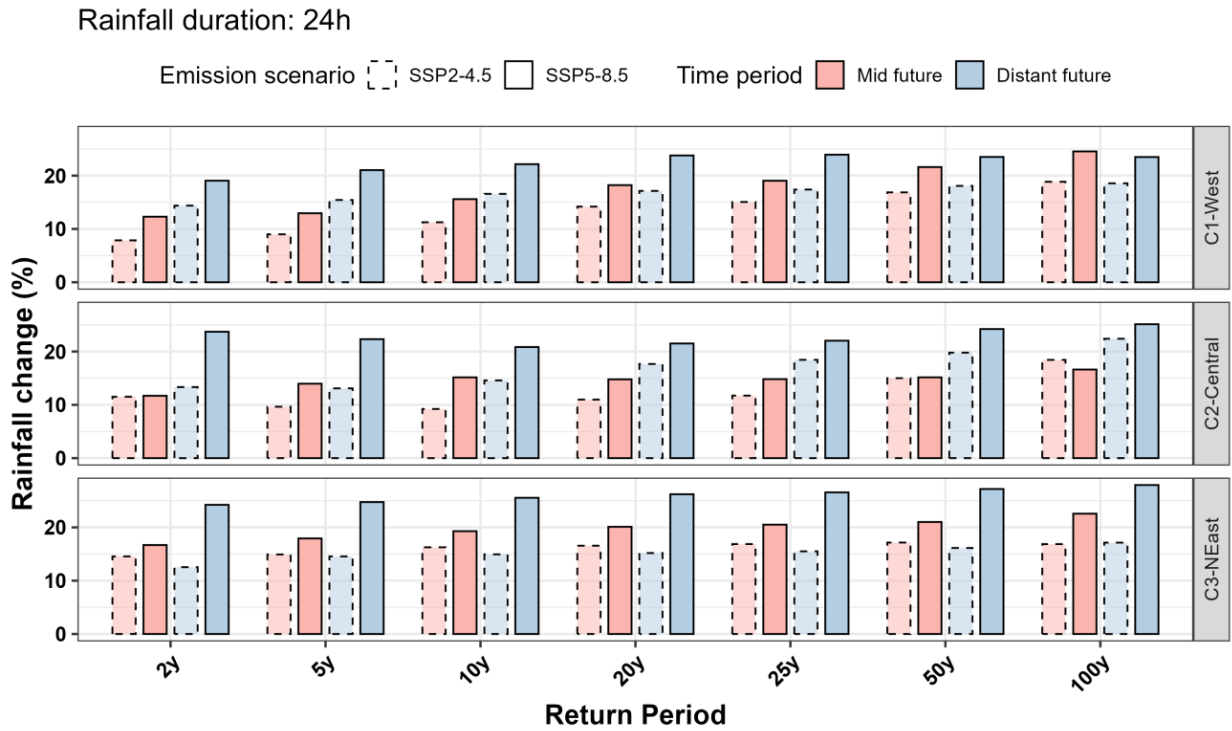

**Figure 17.** As in Figure 16 but for 24-hour duration rainfall.

**Table 6.** Hazard, vulnerability, and risk levels for the selected locations, along with the corresponding risk-based rainfall adjustments. The rainfall increase percentages are based on the SSP5-8.5 scenario, using the maximum value between the mid- and far-future projections.

| Drainage location | Hazard level | SVI level | Transportation vulnerability | Environment vulnerability | Risk level | Rainfall increase% (maximum) | Risk-based adjustment |
|---|---|---|---|---|---|---|---|
| C1-West | 4 | 3 | 4 | 2 | 3 | 27 | 16.2 |
| C2-Central | 3 | 5 | 4 | 2 | 3 | 25 | 15 |
| C3-NEast | 3 | 2 | 4 | 4 | 3 | 32 | 19.2 |

## 8. Concluding remarks

Increased rainfall intensity tends to raise the risk of flooding, which necessitates more robust and adaptive drainage systems. However, the lack of a standardized approach for integrating future climate projections into infrastructure design complicates decision-making. Existing methodologies often rely on historical trends or generic climate change adjustments that fail to capture the localized and intensified flood risks posed by shifting precipitation patterns. Moving away from a static, trend-based climate adjustment approach toward a dynamic, risk-driven methodology, this paper introduces a structured methodology that integrates physiographic and meteorologic hazards with vulnerability assessments to guide infrastructure design. This approach takes climate change into consideration, but with a targeted risk-based adjustment of future rainfall projections. The framework is designed for broad applicability across diverse climatic, geographic, and administrative settings. Importantly, the framework relies on readily available data, which enhances its feasibility and adoption in regions with limited data resources. Linear adjustment is used to scale projected IDF values based on risk levels. However, non-linear climate trends are accounted for by evaluating projections across multiple future periods and using the maximum projected increase in the meteorological hazard calculation. Designers are also provided with projected increases for both mid- and far-future scenarios to ensure conservative and informed selection.

We demonstrated that applying the maximum projected rainfall increase uniformly, without accounting for varying risk levels and inherent uncertainties in future rainfall projections, is not always a practical solution. Therefore, designs can be modified based on localized risk assessments, which allows resources to be directed where they are most needed. Incorporating a risk-informed approach enables engineers to refine IDF curves based on actual, localized risk, leading to drainage systems that are robust enough for future conditions without costly and unnecessary overdesign. Incorporating expert judgment and real-world infrastructure conditions can strengthen this approach. The hierarchical framework introduced in this study is designed to accommodate more formal methods such as the Analytic Hierarchy Process (AHP) in future applications, should expert-based pairwise comparison data become available (Al-Omari et al., 2024; Tabasi et al., 2025). Given the inherent subjectivity in weighting hazard and vulnerability components, the current approach uses scenario-based sensitivity analysis to assess the impact of varying weights. Future work will focus on integrating empirical calibration and participatory methods to refine these parameters, thereby improving the reliability and practical aspects of risk assessments. Data-driven approaches like statistical clustering (Ma et al., 2021) can help identify regional patterns to support region-specific weight assignments. Machine learning could also be used to optimize or automate weighting based on historical impacts or expert input (Bui et al., 2023; Fereshtehpour et al., 2024). The effectiveness of such approaches would depend on the availability and quality of relevant datasets (Te et al., 2024). Even as methods for risk assessment and rainfall projection evolve, adopting the proposed risk-informed framework in this study within official guidelines will allow transportation agencies to update drainage infrastructure proactively, align investment with actual threat levels, and systematically embed climate resilience into future road projects.

**Acknowledgement**

This research was supported by a grant from MTO under their Highway Infrastructure Innovations Funding Program (HIIFP), and we thank them for both their financial and technical support. Opinions expressed in this paper are those of the authors and may not necessarily reflect the views and policies of MTO. We thank Negin Binesh for performing the calculations of future projections of extreme precipitation from the IDF_CC tool.

**CRediT authorship contribution statement**

**Mohammad Fereshtehpour:** Writing – original draft, Visualization, Methodology, Investigation, Formal analysis, Data curation, Conceptualization. **Rashid Nashir:** Writing – review & editing, Conceptualization, Resources, Supervision, Project administration, Funding acquisition. **Neil F. Tandon:** Writing – review & editing, Conceptualization, Validation, Conceptualization, Supervision, Project administration, Funding acquisition.

**Declaration of competing interest**

The authors declare that they have no known competing financial interests or personal relationships that could have appeared to influence the work reported in this paper.

**Data availability**

The datasets used in this paper have been cited with complete information provided in the references. Codes used to generate the figures are available from the authors upon request.